\documentclass[12pt]{article}
\usepackage{amsmath}

\input{tcilatex}

\begin{document}

\title{{\LARGE STANDARD AND EMBEDDED SOLITONS IN NEMATIC OPTICAL FIBERS}}
\author{R. F. Rodr\'{\i}guez$^{1,3}$\thanks{
Fellow of SNI, Mexico} \thanks{
Correspondence author. E-mail: zepeda@fisica.unam.mx}, J. A. Reyes$^{1,3}$ 
\thanks{
Fellow of SNI, Mexico}$,$ \and A. Espinosa-Cer\'{o}n$^{4},$ J. Fujioka$%
^{2,3} $ and B.A. Malomed$^{5}$ \\
$^{1}$\emph{Departamento de F\'{\i}sica Qu\'{\i}mica}\\
\emph{\ }$^{2}$\emph{Departamento de Materia Condensada}\\
$^{3}$\emph{Instituto de F\'{\i}sica. Universidad Nacional }\\
\emph{Aut\'{o}noma de} \emph{M\'{e}xico. }\\
\emph{Apdo. Postal 20-364}\\
\emph{01000 M\'{e}xico, D. F., M\'{e}xico} \\
$^{4}$\emph{Facultad de Ciencias, UAEMEX, Toluca 50000,}\\
\emph{\ Edo. de M\'{e}xico, M\'{e}xico}\\
$^{5}\emph{Department}$ \emph{of Interdisciplinary Studies, }\\
\ \emph{Faculty of Engineering, Tel Aviv University, }\\
\ \emph{Tel Aviv 69978, Israel}\\
\\
}
\date{}
\maketitle

\begin{abstract}
A model for a non-Kerr cylindrical nematic fiber is presented. We use the
multiple scales method to show the possibility of constructing different
kinds of wavepackets of transverse magnetic ($TM$) modes propagating through
the fiber. This procedure allows us to generate different hierarchies of
nonlinear partial differential equations ($PDEs$) which describe the
propagation of optical pulses along the fiber. We go beyond the usual weakly
nonlinear limit of a Kerr medium and derive a complex modified Korteweg-de
Vries equation (cmKdV) which governs the dynamics for the amplitude of the
wavepacket. In this derivation the dispersion, self-focussing and
diffraction in the nematic are taken into account. It is shown that this
cmKdV equation has two-parameter families of bright and dark complex
solitons. We show analytically that under certain conditions, the bright
solitons are actually double embedded solitons. We explain why these
solitons do not radiate at all, even though their wavenumbers are contained
in the linear spectrum of the system. We study (numerically and
analytically) the stability of these solitons. Our results show that these
embedded solitons are stable solutions, which is an interesting property
since in most systems the embedded solitons are weakly unstable solutions.
Finally, we close the paper by making comments on the advantages as well as
the limitations of our approach, and on further generalizations of the model
and method presented.
\end{abstract}

\baselineskip=.5cm


\newpage

\section{Introduction}

Theoretical studies on the existence of solitons in liquid crystals ($LCs$)
started in the late sixties and early seventies \cite{lam}-\cite{brochard},
and experimental confirmations were reported subsequently \cite{leger}-\cite
{ribota}. In the case of static solitons in LCs, the molecular
configurations may be obtained from the Lagrange equations derived from the
Helmholtz free energy, whereas for propagating solitons the continuous
change in these configurations makes it necessary to take into account the
damping of the molecular motion. For liquid-crystal waveguides, the
nonlinearity necessary for the existence of solitons is provided by the
coupling with the optical field.

Coupling of the dynamics of the velocity and director fields in $LCs$ to
external optical fields renders the relevant dynamical equations highly
nonlinear, which makes it possible to have solitary waves of the director
field with or without involving the fluid motion. Furthermore, the strong
coupling of the director to light makes any director-wave more easily
detectable by optical methods than it is in isotropic fluids, where only the
flow field is observable.

Some nonlinear partial differential equations ($PDEs$) appearing in the
liquid-crystal theory give rise to exact soliton solutions. These are the
Korteweg-de Vries ($KdV$), nonlinear Schr\"{o}dinger ($NLS$), and the
sine-Gordon ($sG$) equations \cite{bullough}. The $KdV$ equation describes a
medium with weak nonlinearity and weak dispersion, whereas the $NLS$
equation describes situations where weak nonlinearity and strong dispersion
prevail, such as the propagation of signals in liquid-crystal optical fibers.

Passing continuous laser beams through nematic LCs reveals the existence of
static spatial patterns in cylindrical \cite{braun1} and planar \cite{braun2}
geometries. The basic physical mechanism which support these
time-independent patterns is the balance between the nonlinear refraction
(self-focussing) and spatial diffraction in the nematic. However, when the
propagation of wavepackets, rather than continuous beams, is considered, a
different situation occurs. The envelope of the wavepacket obeys an $NLS$
equation, which takes into account self-focussing, dispersion and
diffraction in the nematic \cite{Rodriguez1}, \cite{adrian1}, \cite
{rodriguez2}. This equation has soliton solutions whose speed, time and
length scales may be estimated by using experimentally measured values of
the corresponding parameters \cite{chinos}. However, the usual analysis of
this situation is based on the assumption that the $LC$ behaves as a Kerr
medium and that, consequently, strong dispersion and weak nonlinearity, at
order $O(q^{3})$, with respect the field amplitude $q$, should be taken into
account As it will be discussed below, $q$ measures the ratio of the
electric-field energy density and the elastic-energy density of the nematic
and it is, therefore, a measure of the coupling between the optical field
and the fluid. However, although truncating the analysis at the $O(q^{3})$
order may be a very reasonable assumption for solid-state optical media, the
soft nature of the LCs suggest that the neglect of higher-order
contributions may not necessarily be a good assumption in this case.

Recently, the formation of spatial solitary waves in nematic LCs with at the
light-power level of a few milliwatts has attracted a good deal of interest 
\cite{buryak}, \cite{assanto}, \cite{peccianti2}, \cite{karpierz}. It has
been experimentally shown that the nonlinearity of these media can support
solitons in $LC$ line waveguides \cite{warenhem}, \cite{peccianti1}.

The main purpose of the present work is to develop an approach that allows
to generate $PDEs$ which describe the propagation of optical pulses in
nematic $LC$ waveguides beyond the weakly nonlinear limit corresponding to
the Kerr medium. More specifically, we show that to $O(q^{4})$, and assuming
that attenuation effects are small, the evolution of the amplitude of
propagating transverse-magnetic transverse-magnetic ($TM$) modes is governed
by an equation\ with a derivative nonlinearity, which is the complex
modified $KdV$ ($cmKdV$) equation, 
\begin{equation}
u_{z}-\varepsilon \,u_{ttt}-\gamma \,\left| u\right| ^{2}u_{t}=0.
\label{1.1}
\end{equation}
see Eq. (\ref{4aa.1}) below.

The paper is organized as follows. In Sec. 2 we introduce a model of a
cylindrical nematic cell and set up basic coupled equations for the
orientational and optical fields. We formulate an iterative procedure to
expand these equations in terms of the coupling parameter $q$, which leads
to a specific hierarchy of $PDEs$. Then, in Sec. 3 we derive dynamical
equations governing the evolution of the amplitude of propagating $TM$ modes
up to the order $O(q^{4})$. Rescaling the equations, we show that the
standard $NLS$ equation is obtained at order $O(q^{3})$, and that the
equation corresponding to $O(q^{4})$ is indeed the $cmKdV$ equation (\ref
{1.1}). In Sec. 4, soliton solutions to this equation are studied. In
particular, it is shown that the equation has ordinary bright- and
dark-soliton solutions, and a continuous family of \textit{embedded solitons 
}($ESs$), i.e. solitary waves which exist inside the system's continuous
spectrum of linear waves \cite{yang1}\textit{. }In Sec. 5 we discuss why the 
$ES$s can exist in Eq. (\ref{1.1}) without emitting any radiation, even
though their wave numbers belong to the linear spectrum. In Sec. 6 we study
the stability of the ESs.\ \ We conclude the paper in Sec. 7, which
summarizes the results and compares them to previously published ones. We
also point out advantages and limitations of our approach, and discuss
possible ways to generalize it.

\section{The Model and Basic Equations}

We consider a cylindrical waveguide with an isotropic core of radius $a$,
dielectric constant $\epsilon _{c}$ and a quiescent nematic $LC$ cladding of
radius $b$. The initial orientational state is depicted in Fig. 1, where the
director field obeys the following axial strong-anchoring boundary
conditions, 
\begin{equation}
\hat{n}(r=a,z)=\hat{n}(r=b,z)=\widehat{e}_{z}.  \label{2.1}
\end{equation}
An optical beam is launched into the guide and propagates through the $LC$.
If the field is strong enough to exceed the orientation-transition
threshold, the initial configuration is changed by reorienting the director
field. We assume that the induced reorientation occurs only occurs in the $%
\left( r,z\right) $ plane, so that

\begin{equation}
\hat{n}(r,z)=\sin \theta \widehat{e}_{r}+\cos \theta \widehat{e}_{z},
\label{2.2}
\end{equation}
where $\widehat{e}_{r}$ and $\widehat{e}_{z}$ are the unit vectors of the
cylindric coordinates.

Although the incident beam is neither planar nor Gaussian, the normal modes
within the cavity are cylindrical plane waves propagating along the $z$
axis. In previous works it has been shown that only the $TM$ modes, with
nonzero components $E_{r}(r,z,t)$, $E_{z}(r,z,t)$ and $H_{\phi }(r,z,t)$ of
the electromagnetic field, couple to the reorientation dynamics of the
director field \cite{adrian1}, \cite{Rodriguez1}, \cite{pulsed}. As it can
be shown that $E_{r}(r,z,t)$ and $E_{z}(r,z,t)$ may be expressed in terms of 
$H_{\phi }(r,z,t)$, below we only describe the dynamics of the component $%
H_{\phi }(r,z,t)$. The relevant dynamical equations, which take into account
retardation effects, are given by Eqs. (8) and (9) of Ref. \cite{pulsed},
namely,

\begin{eqnarray}
&&\frac{\partial ^{2}\theta }{\partial \zeta ^{2}}+\frac{1}{x}\frac{\partial 
}{\partial r}\left( x\frac{\partial \theta }{\partial x}\right) -\frac{\sin
\theta \cos \theta }{x^{2}}  \notag \\
&&-q^{2}\left[ \frac{\cos 2\theta }{x}\left( \mathcal{E}_{z}^{\ast
}\int^{t}dt^{\prime }\frac{\partial x\mathcal{H}_{\phi }}{\partial x}+ 
\mathcal{E}_{r}^{\ast }\int^{t}dt^{\prime }\frac{\partial \mathcal{H}_{\phi
} }{\partial \zeta }\right) +\right.  \notag \\
&&\left. \frac{\sin 2\theta }{x^{2}}\left( -x\mathcal{E}_{r}^{\ast
}\int^{t}dt^{\prime }\frac{\partial \mathcal{H}_{\phi }}{\partial \zeta }+ 
\mathcal{E}_{z}^{\ast }\int^{t}dt^{\prime }\frac{\partial x\mathcal{H}_{\phi
}}{\partial x}\right) \right] =0,  \label{2.3}
\end{eqnarray}

\begin{eqnarray}
\frac{a^{2}}{c^{2}}\frac{\partial ^{2}H_{\phi }}{\partial t^{2}} &=&-\int
dt^{^{\prime }}\frac{\left( \frac{\partial ^{2}H_{\phi }}{\partial \zeta
^{2} }+\frac{\partial ^{2}H_{\phi }}{\partial x^{2}}\right) \left(
t-t^{^{\prime }}\right) }{\epsilon _{\perp }\left( \overrightarrow{r^{\prime
}},t^{\prime }\right) }+\frac{\partial ^{2}}{\partial t\partial \zeta }\int
dt^{^{\prime }}  \notag \\
&&\frac{\epsilon _{a}}{\epsilon _{\perp }\epsilon _{\parallel }}\left(
t^{\prime }\right) \left[ -\sin ^{2}\theta \frac{\partial H_{\phi }} {%
\partial \zeta }+\sin \theta \cos \theta \frac{1}{x}\frac{\partial } {%
\partial x}xH_{\phi }\right] \left( t-t^{^{\prime }}\right)  \notag \\
&&-\frac{\partial ^{2}}{\partial t\partial x}\int dt^{^{\prime }}\frac{
\epsilon _{a}}{\epsilon _{\perp }\epsilon _{\parallel }}\left( t^{\prime
}\right) \left[ -\sin \theta \cos \theta \frac{\partial H_{\phi }}{\partial
\zeta }+\right.  \notag \\
&&\left. \cos ^{2}\theta \frac{1}{x}\frac{\partial }{\partial x}xH_{\phi} %
\right] \left( t-t^{^{\prime }}\right) ,  \label{2.4}
\end{eqnarray}
with

\begin{equation}
\overset{\rightarrow }{\mathcal{E}}\left( \overrightarrow{r},t\right) = 
\frac{1}{\epsilon _{0}}\int dt^{^{\prime }}\int^{t}dt^{"}\frac{\epsilon _{a} 
}{\epsilon _{\perp }\epsilon _{\parallel }}\left( t^{"}-t^{^{\prime
}}\right) \hat{n}\hat{n}\cdot \nabla \times \overrightarrow{\mathcal{H}}
\left( \overrightarrow{r^{\prime }},t^{\prime }\right) .  \label{2.5}
\end{equation}
In these equations, we have used dimensionless variables, $\zeta \equiv z/a$%
, $x\equiv r/a,$ $H_{\phi }\equiv \mathcal{H}_{\phi }/(c\epsilon _{0}E_{0}),$
$E_{i}^{a}\equiv \mathcal{E}_{i}^{a}/E_{0}$ , $i=r,z,$ where $E_{0}$ is the
amplitude of the incident field. The speed of light in vacuum is $c=1/\sqrt{
\mu _{0}\epsilon _{0}}$ , where $\mu _{0}$ and $\epsilon _{0}$ are,
respectively, the magnetic permeability and electric permittivity of free
space. The dielectric anisotropy of the nematic, $\epsilon _{a}\equiv
\epsilon _{\parallel }-\epsilon _{\perp }$, is defined in terms of the
dielectric constant for directions parallel ($\epsilon _{\parallel }$) and
perpendicular ($\epsilon _{\perp }$) to the director. As mentioned in Sec.
1, $q^{2}\equiv \epsilon _{0}E_{0}^{2}a^{2}/K$ is the dimensionless ratio
between the electric-field energy density and the elastic-energy density of
the nematic, where $K$ is its elastic constant in the equal constants
approximation. Thus, $q^{2}$ is a measure of the coupling between the
optical field and the $LC$. We stress that, in writing Eqs. (\ref{2.3}) and (%
\ref{2.4}), large difference between the time scales of slow reorientation
dynamics and rapid variations of the electromagnetic field was explicitly
taken into account, and as a consequence the time derivatives of $\theta $
were ignored.

When the coupling between the $TM$ \ mode $H_{\phi }(r,z,t)$ and the
reorientation field $\theta (r,z,t)$ is negligible ($q=0$), the propagating
modes are represented by quasi-planar waves. However, if the nonlinearities
in Eq. (\ref{2.3}) are taken into account by considering finite $q$, they
cause space and time variations of the field $H_{\phi }(r,z,t)$, due to
generation of higher-order harmonics which feedback to the original modes.

We assume that the interaction between the optical field and the
reorientation in the nematic is stronger than in the weakly nonlinear limit
(Kerr medium) which corresponds to $q=1$ \cite{Rodriguez1}. Furthermore, in
all the analysis we neglect all the backflow effects associated with the
reorientation or caused by external flows \cite{Rodriguez1prima}. Thus, we
solve the coupled equations (\ref{2.3}) and (\ref{2.4}) by assuming the
following coupled expansions of $\theta $ and $H_{\phi }$ in powers of $q$, 
\begin{equation}
\theta =\theta ^{(o)}+q^{2}\left| A(\Xi ,T)U(x,\omega )\right| ^{2}\theta
^{(1)}(x)+q^{4}\left| A(\Xi ,T)U(x,\omega )\right| ^{4}\theta ^{(2)}(x)+...,
\label{2.6}
\end{equation}

\begin{eqnarray}
H_{\phi }\left( x,\zeta ,t\right) &=&qU_{\phi }\left( x,\omega _{0}+i\lambda 
\frac{\partial }{\partial T}\right) A(\Xi ,T)+q^{2}U^{(2)}+q^{3}U^{(3)} 
\notag \\
&&+q^{4}U^{(4)}+q^{5}U^{(5)}+\mathrm{c.c.}+...,  \label{2.7}
\end{eqnarray}
where $\mathrm{c.c.}$ stands for the complex conjugate.

The rationale behind this assumption is the following. As indicated in Eq. (%
\ref{2.3}), the lowest-order coupling between $\theta $ and $H_{\phi }$
occurs at order $q^{2}$, and it is therefore reasonable to expect that
higher-order terms will also be even in $q$. The fields $\theta ^{(n)}$ with 
$n=0$, $1$, $2..$\ are contributions to $\theta $ at order $n$ , which
satisfy the same hard-anchoring homeotropic boundary conditions as were
given above by Eq. (\ref{2.1}), $\theta (x=1)=\theta (x=b/a)=0$. As usual,
the amplitude $A(\Xi ,T)$ in Eqs. (\ref{2.6}) and (\ref{2.7}), which
represents an envelope of a narrow wavepacket of width $\lambda \equiv
(\omega -\omega _{0})/\omega _{0}$, whose central frequency is $\omega _{0}$%
, is assumed to be a slowly varying function of the variables $\Xi \equiv
\lambda \zeta $ and $T\equiv \lambda t$ . Here $\lambda $ is a small
parameter which measures the dispersion of the wavepacket. In Eqs. (\ref{2.6}%
) and (\ref{2.7}), $U_{\phi }\left( x,\omega _{0}\right) $ is the well-known
linear solution for $H_{\phi }$ which is given explicitly by \cite{jackson}

\begin{equation}
U_{\phi }(x,\omega _{0})=J_{1}^{2}\left( \sqrt{\epsilon _{c}(\frac{\omega
_{0}a}{c})^{2}-\beta ^{2}a^{2}}\right) \sqrt{\frac{\pi }{2\gamma ax}}\exp
(-i\beta a\zeta -\gamma ax),  \label{2.8}
\end{equation}
with $\gamma =\sqrt{\epsilon _{\parallel }\left( \beta ^{2}/\epsilon _{\perp
}-\left( \omega _{0}/c\right) ^{2}\right) }$. Here $J_{1}(x)$ is the Bessel
function of order $1$, and $\beta $ is the propagation constant, which only
takes allowed values calculated in Ref. \cite{pulsed}. The terms
proportional to $U^{(n)}$, $n=2,3,...$ in (\ref{2.7}) are contributions to
the $TM$ modes from to the higher-order optical harmonics that are generated
by the nonlinearities in Eqs. (\ref{2.4}) and (\ref{2.3}).

Note, however, that the relation between the parameters $q$ and $\lambda $
is not unique. For instance, when the wavepacket is very narrow, this
relation is $\lambda =q$ and up to $O(q^{3})$, the expansion leads to the
standard nonlinear Schr\"{o}dinger ($NLS$) equation for $A(\Xi ,T)$ (which
corresponds to the Kerr medium) \cite{Newell}, \cite{Rodriguez1}, \cite
{pulsed}. Therefore the model may be generalized in various ways. Since $q$
and $\lambda $ are small parameters, we assume that $\lambda \equiv
q^{\alpha }$ with some positive $\alpha $. Then $\alpha =1/2$ represents a
wider and $\alpha =2$ a narrower wavepacket. Note that the presence of
higher powers of $q$ implies that these higher-order contributions are
smaller than the dominant term in (\ref{2.6}), which describes a
small-amplitude narrow wavepacket.

Inserting the expression\ (\ref{2.6}) into Eq.(\ref{2.4}) and expanding in
powers of \ $q$, it is straightforward to rewrite Eq.(\ref{2.4}) as

\begin{equation}
\widehat{L}(\beta ,\omega ,x)H_{\phi }+q^{2}\widehat{F}(H_{\phi })+q^{4} 
\widehat{G}(H_{\phi })=0,  \label{2.9}
\end{equation}
where the linear $\widehat{L}$ operator and nonlinear ones, $\widehat{F}$
and $\widehat{G}$, are defined, respectively, as

\begin{equation}
\widehat{L}\equiv \frac{1}{x^{2}\epsilon _{\perp }\epsilon _{\parallel }} %
\left[ -\epsilon _{\perp }+x^{2}\epsilon _{\parallel }\left( \epsilon
_{\perp }\left( \frac{\omega _{0}}{c}a\right) ^{2}-(\beta a)^{2}\right)
+x\epsilon _{\perp }\frac{\partial }{\partial x}+x^{2}\epsilon _{\perp } 
\frac{\partial ^{2}}{\partial x^{2}}\right] ,  \label{2.10}
\end{equation}

\begin{equation}
\widehat{F}\equiv \frac{\epsilon _{a}\left| A(\zeta )U_{\phi }(x,\omega
)\right| ^{2}}{x\epsilon _{\perp }\epsilon _{\parallel }}i\beta a\left(
U_{\phi }\theta ^{(1)}(x)+3x\theta ^{(1)}(x)\frac{dU_{\phi }}{dx}+U_{\phi }x 
\frac{d\theta ^{(1)}(x)}{dx}\right) A.  \label{2.11}
\end{equation}

\begin{eqnarray}
\widehat{G} &=&-\frac{\epsilon _{a}\left| A(\zeta )U_{\phi }(x,\omega
)\right| ^{4}A(\zeta )}{x^{2}\epsilon _{\perp }\epsilon _{\parallel }}\left[
\left\{ (x^{2}\beta ^{2}-1)\left( \theta ^{(1)}(x)\right) ^{2}-ix\beta
\theta ^{(2)}(x)\right. \right.  \notag \\
&&\left. +2x\theta ^{(1)}(x)\frac{d\theta ^{(1)}(x)}{dx}-ix^{2}\beta \frac{
d\theta ^{(2)}(x)}{dx}\right\} U_{\phi }(x,\omega )+\left\{ x\left( \theta
^{(1)}(x)\right) ^{2}\right.  \notag \\
&&\left. -6ix^{2}\beta \theta ^{(2)}(x)+2x^{2}\theta ^{(1)}(x)\frac{d\theta
^{(1)}(x)}{dx}\right\} \frac{dU_{\phi }(x,\omega )}{dx}  \notag \\
&&+x^{2}\left( \theta ^{(1)}(x)\right) ^{2}\frac{d^{2}U_{\phi }(x,\omega )} {%
dx^{2}}  \notag \\
&&\left. -6ix^{2}\beta \theta ^{(2)}(x)\right] .  \label{2.12}
\end{eqnarray}
Zero-order solutions for the orientation, with $\theta ^{(0)}=0$, and
first-order ones, which gives rise to $\theta ^{(1)}(x)$, were found in \
Ref. \cite{pulsed} by inserting expressions (\ref{2.6}) and (\ref{2.7}) into
Eqs. (\ref{2.3}), (\ref{2.9}) and solving the resulting equations. $\ $This
way, $\theta ^{(1)}(x)$ turned out to be

\begin{eqnarray}
\theta ^{(1)}(x) &=&\frac{\beta a\epsilon _{a}J_{1}^{2}[\sqrt{\epsilon _{c} (%
\frac{\omega _{0}a}{c})^{2}-\beta ^{2}a^{2}]}}{\pi \epsilon _{\perp
}\epsilon _{\parallel }x\left( a^{2}-b^{2}\right) }\left\{
(a^{2}-b^{2})e^{\gamma a(1-x)}\right.  \notag \\
&&+\left. (b^{2}-x^{2}a^{2})+e^{\gamma (a-b)}a^{2}(1-x^{2})\right\} .
\label{2.13}
\end{eqnarray}

To study the dynamics beyond the Kerr approximation, we need to calculate
the fourth-order terms in Eq. (\ref{2.6}), that is, $\theta ^{(2)}(x)$. To
this end,we insert Eqs. (\ref{2.8}), (\ref{2.6}) into Eq. (\ref{2.3}) and
expand the result in powers of $q$ up to the fourth order. This leads to

\begin{eqnarray}
&&\epsilon _{a}\gamma a\frac{\left( 4x^{2}\left[ \epsilon _{\perp }\epsilon
_{\parallel }(\frac{\omega _{0}a}{c})^{2}-\beta ^{2}a^{2}\right] -\epsilon
_{\perp }\right) }{2\pi x^{2}\epsilon _{\perp }\epsilon _{\parallel
}^{2}\left( \beta ^{2}a^{2}-\epsilon _{\perp } (\frac{\omega _{0}a}{c}%
)^{2}\right) }\theta ^{(1)}(x)+  \notag \\
&&\frac{\theta ^{(2)}(x)}{x}-\frac{d\theta ^{(2)}(x)}{dx}-x\frac{d^{2}\theta
^{(2)}(x)}{dx^{2}}=0.  \label{2.14}
\end{eqnarray}
After substituting $\theta ^{(1)}(x)$ from Eq. (\ref{2.13}), this equation
takes the form

\begin{eqnarray}
&&\frac{\beta \epsilon _{a}^{2}\gamma a^{2}}{2\pi ^{2}\epsilon _{\perp
}^{2}\epsilon _{\parallel }^{3}x^{3}\left( a^{2}-b^{2}\right) }\frac{\left(
4x^{2}\left[ \epsilon _{\perp }\epsilon _{\parallel }(\frac{\omega _{0}a}{c}
)^{2}-\beta ^{2}a^{2}\right] -\epsilon _{\perp }\right) }{\left( \beta
^{2}a^{2}-\epsilon _{\perp }(\frac{\omega _{0}a}{c})^{2}\right) }  \notag \\
&&J_{1}^{2}[\sqrt{\epsilon _{c}(\frac{\omega _{0}a}{c})^{2}-\beta ^{2}a^{2}}]
\notag \\
&&\left\{ (a^{2}-b^{2})e^{\gamma a(1-x)}+(b^{2}-x^{2}a^{2})+e^{\gamma
(a-b)}a^{2}(1-x^{2})\right\}  \notag \\
&&+\frac{\theta ^{(2)}(x)}{x}-\frac{d\theta ^{(2)}(x)}{dx}-x\frac{
d^{2}\theta ^{(2)}(x)}{dx^{2}}=0.  \label{2.15}
\end{eqnarray}
In spite of its apparent complexity, this linear differential equation for $%
\theta ^{(2)}(x)$ can be easily solved by imposing the planar
strong-anchoring boundary conditions for $\theta $, as explained above. The
solution can then be written in terms of the exponential-integral function,
and if the resulting expressions are approximated by asymptotic expressions
for this function, we obtain

\begin{eqnarray}
\theta ^{(2)}(x) &=&\frac{\beta \epsilon _{a}e^{-(b+ax)\gamma }J_{1}^{2} [%
\sqrt{\epsilon _{c}(\frac{\omega _{0}a}{c})^{2}-\beta ^{2}a^{2}}]} {%
24x^{2}a^{2}b(a+b)^{2}(b-a)\pi ^{3}\gamma ^{2}\epsilon _{\perp }^{3}\epsilon
_{\parallel }^{2}}  \notag \\
&&\left[ e^{(1+x)a\gamma }[A_{4}x^{4}+A_{3}x^{3}+A_{2}x^{2}\right.  \notag \\
&&+A_{1}x+A_{0}]+e^{-(b+a)\gamma }(B_{1}x+B_{0})]  \notag \\
&&+\left. e^{(b+ax)\gamma }(C_{4}x^{4}+C_{3}x^{3}+C_{2}x^{2}+C_{1}x+C_{0}) 
\right] .  \label{2.16}
\end{eqnarray}
\newline
While this compact form for $\theta ^{(2)}(x)$ is sufficient for our
discussion below, expressions for the coefficients $A_{0}$, $A_{1}$, $A_{2}$%
, $A_{3}$, $A_{4}$, $B_{0}$, $B_{1}$, $C_{0}$, $C_{1}$, $C_{2}$, $C_{3}$ and 
$C_{4}$ that appear in Eq. (\ref{2.16}) are given in Appendix A.

To conclude this section, it is relevant to stress that the above derivation
ignored dissipative loss in the $LC$ medium. In fact, the physical condition
for the applicability of this assumption is that the propagation distance to
be passed by excitations (solitons) is essentially smaller than a
characteristic dissipative-loss length. This condition can be readily met in
situations of physical relevance.

\section{The Envelope Dynamics}

We now aim to derive an equation for the envelope $A(\Xi ,T)$ by dint of the
same procedure that was used in Ref. \cite{Rodriguez1} for the weakly
nonlinear case. To this end, we substitute Eq. (\ref{2.7}) into Eq. (\ref
{2.9}), and identify the Fourier variables 
\begin{equation}
i\beta a=i\beta _{0}a+q^{\alpha }\frac{\partial }{\partial \Xi _{1}}
+q^{2\alpha }\frac{\partial }{\partial \Xi _{2}}+q^{3\alpha }\frac{\partial 
}{\partial \Xi _{3}}+q^{4\alpha }\frac{\partial }{\partial \Xi _{4}}\,,
\label{3.1}
\end{equation}
\begin{equation}
-i\omega =-i\omega _{0}+q^{\alpha }\frac{\partial }{\partial T},  \label{3.2}
\end{equation}
where the variables $\Xi _{n},n=1,2,3,4,$ are related to the spatial scales
associated with upper harmonics contributions, that is, $Z\equiv q^{n\alpha
}\Xi _{n}$. This substitution leads to an equation,

\begin{eqnarray}
0 &=&\widehat{L}[i\beta _{0}a+q^{\alpha }\frac{\partial }{\partial \Xi _{1}}
+q^{2\alpha }\frac{\partial }{\partial \Xi _{2}}+q^{3\alpha }\frac{\partial 
}{\partial \Xi _{3}}+q^{4\alpha }\frac{\partial }{\partial \Xi _{4}}
,-i\omega _{0} \\
&&+q^{\alpha }\frac{\partial }{\partial T}]H_{\phi }(x,\zeta ,t)+q^{2} 
\widehat{F}[H_{\phi }\left( x,\zeta ,t\right) ]+q^{4}\widehat{G[}(H_{\phi
}\left( x,\zeta ,t\right) ].  \label{3.3}
\end{eqnarray}

We now fix $\alpha =1$, which means selection of the type of the wave packet
to be considered; choosing $\alpha =2$ or $\alpha =1/2$ would imply,
respectively, a narrower or wider packet of the $TM$ modes than for $\alpha
=1$. In this case, we collect contributions to the same power of $q$,
arriving at the expressions

\begin{equation}
q^{1}:\widehat{L}\left( i\beta _{0}a,-i\omega _{0},x\right) U_{\phi }\left(
x,\omega _{0}\right) A=0,  \label{3.4}
\end{equation}

\begin{equation}
q^{2}:\widehat{L}U^{(1)}=\left( i\widehat{L}\left( i\beta _{0}a,-i\omega
_{0}\right) \frac{\partial U_{\phi }}{\partial \omega }\frac{\partial } {%
\partial T}+U_{\phi }\widehat{L}_{2}\frac{\partial }{\partial T}+\widehat{L}
_{1}U_{\phi }\frac{\partial }{\partial \Xi _{1}}\right) A,  \label{3.5}
\end{equation}

\begin{equation}
q^{3}:\widehat{L}U^{(2)}=\hat{S}_{2}\left( \frac{\partial ^{2}U_{\phi }} {%
\partial \omega ^{2}},\frac{\partial U_{\phi }}{\partial \omega },U_{\phi
}\right) A-\widehat{F}(U_{\phi }\left( x,\omega _{0}\right) A),  \label{3.6}
\end{equation}

\begin{eqnarray}
q^{4} &:&\widehat{L}U^{(3)}=\hat{S}_{3}\left( \frac{\partial ^{3}U_{\phi }} {%
\partial \omega ^{3}},\frac{\partial ^{2}U_{\phi }}{\partial \omega ^{2}}, 
\frac{\partial U_{\phi }}{\partial \omega },U_{\phi }\right) A+  \notag \\
&&R_{3}\left( \frac{\partial U_{\phi }}{\partial \omega },\frac{\partial
U_{\phi }}{\partial x},\frac{\partial ^{2}U_{\phi }}{\partial \omega
\partial x},\theta ^{(1)}(x),\frac{d\theta ^{(1)}(x)}{dx}\right) \left|
A\right| ^{2}A,  \label{3.7}
\end{eqnarray}
where $\widehat{L}_{n},$ $n=1,2,$ denotes the derivative of $\ \widehat{L}
\left( i\beta _{0}a,-i\omega _{0}\right) $ with respect to its first or
second argument. Clearly, the same procedure can be carried out for $\alpha
=1/2$ or $2$.

Note that Eq. (\ref{3.4}) is actually the usual dispersion relation $%
\widehat{L}\left( i\beta _{0}a,-i\omega _{0}\right) $ $U_{\phi }\left(
x,\omega _{0}\right) =0$, which confirms our approximation, since $H_{\phi
}\left( x,\omega _{0}\right) $ already satisfies this equation to the first
order in $q$. To simplify Eqs. (\ref{3.5}) -- (\ref{3.7}) we take the first
four derivatives of Eq. (\ref{3.4}) with respect to $\omega $. This leads to
a set of linear inhomogeneous equations for $U^{(n)}$, the existence of
solutions to which is secured by the so-called alternative Fredholm
condition \cite{Zwillinger}. This condition is fulfilled if $\widehat{L}
U_{\phi }\left( x,\omega _{0}\right) =0$ and if $U_{\phi }\left( x,\omega
_{0}\right) \rightarrow 0$ as $x\rightarrow \infty $. In our case, this
reads explicitly 
\begin{equation}
\left\langle \widehat{L}U^{(n)},U_{\phi }\right\rangle
=\int_{1}^{b/a}U_{\phi }\widehat{L}U^{(n)}dx=0,\,n=1,2,3,4.  \label{3.9}
\end{equation}
By applying the relations (\ref{3.9}) to Eqs. (\ref{3.5}) -- (\ref{3.7}),
substituting the four first derivatives of Eq. (\ref{3.4}) into them, and
collecting terms in front of the same power of $q$, we obtain the following
equations for $A(\Xi ,T)$ on each of the spatial scales $\Xi $, $\Xi _{1}$, $%
\Xi _{2}$, $\Xi _{3}$, $\Xi _{4}$, for the successive orders in $q$, 
\begin{equation}
q^{2}:\frac{\partial A}{\partial \Xi _{1}}+a\frac{d\beta }{d\omega }\frac{
\partial A}{\partial T}=0,  \label{3.10}
\end{equation}

\begin{equation}
q^{3}:\frac{\partial A}{\partial \Xi _{2}}+i\frac{d^{2}\beta }{d\omega ^{2}} 
\frac{\partial ^{2}A}{\partial T^{2}}+i\beta n_{2}A\left| A\right| ^{2}=0,
\label{3.11}
\end{equation}

\begin{equation}
q^{4}:\frac{\partial A}{\partial \Xi _{3}}-\frac{1}{6}\frac{d^{3}\beta} {%
d\omega ^{3}}\frac{\partial ^{3}A}{\partial T^{3}}-\beta n_{3}\left|
A\right| ^{2}\frac{\partial A}{\partial T}=0.  \label{3.12}
\end{equation}

Here, dimensionless coefficients $\overline{n}_{2}$ and $\overline{n}_{3}$
are defined as follows: 
\begin{eqnarray}
\overline{n}_{2} &=&\frac{1}{4}\epsilon _{a}^{2}\beta a^{3}J_{1}\left( \frac{
a}{c}\sqrt{\left( \epsilon _{c}\omega _{0}^{2}-\beta ^{2}c^{2}\right) }
\right) ^{4}\allowbreak e^{-\gamma b+2\gamma a}  \notag \\
&&\frac{-ae^{-3\gamma b}+ae^{\gamma \left( a-4b\right) }+be^{-\gamma \left(
4a-b\right) }-be^{-3\gamma a}}{\pi \epsilon _{\parallel }^{2}b\left(
a^{2}-b^{2}\right) \epsilon _{\perp }\left( -e^{-2\gamma b}+e^{-2\gamma
a}\right) },  \label{3.14}
\end{eqnarray}

\begin{eqnarray}
\bar{n}_{3} &=&-\frac{\epsilon _{\perp }\omega _{0}}{4\beta }
\int_{1}^{b/a}\left( i\frac{\overline{n}_{2}}{\epsilon _{\perp }}U_{\phi } %
\left[ \beta \frac{\partial U_{\phi }}{\partial \omega }-U_{\phi }\frac{
d\beta }{d\omega }\right] +\frac{3\beta \epsilon _{a}}{x\epsilon _{\perp
}\epsilon _{\parallel }}\left( U_{\phi }\right) ^{3}\frac{\partial U_{\phi } 
}{\partial \omega }\frac{dx\theta ^{(1)}(x)}{dx}+\right.  \notag \\
&&\left. \frac{2\beta \epsilon _{a}}{x\epsilon _{\perp }\epsilon _{\parallel
}}\theta ^{(1)}(x)\left( U_{\phi }\right) ^{2}\left[ 2\frac{\partial U_{\phi
}}{\partial x}\frac{\partial U_{\phi }}{\partial \omega }+U_{\phi }\frac{
\partial ^{2}U_{\phi }}{\partial x\partial \omega }\right] \right)
dx/\int_{1}^{b/a}\left( U_{\phi }\right) ^{2}dx,  \label{3.15}
\end{eqnarray}
The coefficient $\overline{n}_{2}$ is related with the nonlinear diffraction
index $n_{2}$ through the expression $\overline{n}_{2}\equiv Kn_{2}/\epsilon
_{0}a^{2}$. Similarly, we define a nonlinear diffraction index at the next
order beyond the Kerr approximation by $\overline{n}_{3}\equiv \omega
_{0}Kn_{3}/\epsilon _{0}a^{2}$; it is proportional to the coefficient in
front of the nonlinear term in Eq. (\ref{3.12}).\ Note that Eq. (\ref{3.10})
simply describes a wavepacket in the linear medium, while Eq. (\ref{3.11})
is the well-known $NLS$ equation which gives rise to robust soliton pulses.
The equations corresponding to the orders $q^{2}$ and $q^{3}$ are well-known
ones, and they have also been derived and analyzed in Ref. \cite{pulsed}. In
the next section we focus on Eq. (\ref{3.12}), which was derived at order $%
q^{4}$.

\section{Double Embedded Solitons}

Equation (\ref{3.12}) may be rewritten in a rescaled form by introducing the
dimensionless variables $u\equiv A/A_{0}$, $\xi \equiv \Xi _{4}/Z_{04}$, and 
$\tau \equiv T/T_{04}$, where $Z_{04}$ and $T_{04}$ are space and time
scales, and $A_{0}$ is the initial amplitude of the optical pulse. In terms
of these variables, Eq. (\ref{3.12}) becomes 
\begin{equation}
\frac{\partial u}{\partial \xi }-\varepsilon \,\frac{\partial ^{3}u} {%
\partial \tau ^{3}}-\gamma \left| u\right| ^{2}\frac{\partial u}{\partial
\tau }=0,  \label{4aa.1}
\end{equation}
where we have defined the dimensionless coefficients $\varepsilon $ and $%
\gamma $ as 
\begin{equation}
\varepsilon =\frac{1}{6}\frac{Z_{04}}{T_{04}^{\,\,\,3}}\frac{d^{3}\beta } {%
d\omega ^{3}},  \label{4aa.2}
\end{equation}
\begin{equation}
\gamma =\beta n_{3}A_{0}^{\,2}\frac{Z_{04}}{T_{04}}.  \label{4aa.3}
\end{equation}
In what follows below, we will consider Eq. (\ref{4aa.1}) in the form of Eq.
(\ref{1.1}), i.e., with $\xi $ and $\tau $ replaced by $z$ and $t$.

Equation (\ref{4aa.1}), or equivalently Eq. (\ref{1.1}), reduces to the real
modified Korteweg de Vries ($mKdV$) equation when we restrict $u(z,t)$ to be
real, hence all the real solutions of the $mKdV$ equation, including $N$%
-soliton ones, are also solutions of (\ref{1.1}). On the other hand, Eq. (%
\ref{1.1}) also has complex solutions which include, as it will be discussed
below, two-parameter families of bright and dark complex solitons. Actually,
the existence of these complex solutions of Eq. (\ref{1.1}) was pointed out
by Ablowitz and Segur as early as 1981 \cite{ablowitz}. The precise form of
the bright solitons in the particular case when $\varepsilon =6\gamma $ was
presented recently by Karpman \textit{et al}. \cite{karpman}.

In the general case the bright-soliton solutions to Eq. (\ref{1.1}) may be
found by substituting a straightforward trial function in this equation, 
\begin{equation}
u(z,t)=A\;\mathrm{sech}\left( \frac{t-az}{w}\right) e^{i(qz+rt)}.
\label{4aa.4}
\end{equation}
This substitution shows that (\ref{4aa.4}) is indeed a solution of (\ref{1.1}%
), provided that 
\begin{equation}
A^{2}w^{2}=\frac{6\varepsilon }{\gamma },  \label{4aa.5}
\end{equation}

\begin{equation}
a=3\varepsilon r^{2}-\frac{1}{6}\gamma A^{2},  \label{4aa.6}
\end{equation}

\begin{equation}
q=\frac{1}{2}\gamma A^{2}r-\varepsilon r^{3}.  \label{4aa.7}
\end{equation}
Condition (\ref{4aa.5}) implies that the bright soliton solution (\ref{4aa.4}%
) only exists for $\epsilon \gamma >0$, which implies that, in the opposite
case, the nonlinearity and linear dispersion cannot be in balance. Moreover,
since we have five free parameters in Eq. (\ref{4aa.4}) and only three
conditions (\ref{4aa.5}) -- (\ref{4aa.7}), these expressions define a
two-parameter family of bright soliton solutions of Eq. (\ref{1.1}), so that
the following pairs of the parameters can be chosen arbitrarily: ($A,r$), ($%
w,r$), ($A,q$), or ($w,q$). The family includes, as particular cases, the
real one-soliton solutions of the $mKdV$ equation, which are obtained when $%
r=0$.

In a similar way, dark solitons of Eq. (\ref{1.1}) can be found by
substituting the trial function 
\begin{equation}
u(z,t)=A_{d}\,\tanh \left( \frac{t-a_{d}\,z}{w_{d}}\right)
\,\,e^{i(q_{d}\,z\,+\,r_{d\,}t)}.  \label{jf38}
\end{equation}
This substitution shows that this \textit{ansatz} solves Eq. (\ref{1.1}) if
the following conditions are satisfied 
\begin{equation}
A_{d}^{\;2}\,w_{d}^{\;2}=-\frac{6\,\varepsilon }{\gamma },  \label{jf39}
\end{equation}
\begin{equation}
a_{d}=3\,\varepsilon \,r_{d}^{\;2}-\frac{1}{3}\,\gamma \,A_{d}^{\;2},
\label{jf40}
\end{equation}
\begin{equation}
q_{d}=\gamma \,A_{d}^{\;2}r_{d}-\varepsilon \,r_{d}^{\;3},  \label{jf41}
\end{equation}
which are similar to the conditions (\ref{4aa.5}) -- (\ref{4aa.7}) for the
bright solitons. As in the bright case, the conditions (\ref{jf39}) -- (\ref
{jf41}) permit us to choose freely any of the following pairs of parameters: 
$(A,r)$, $(w,r)$, $(A,q)$, or $(w,q)$. Thus, Eqs. (\ref{jf38})-(\ref{jf41})
define a two-parameter family of dark-soliton solutions of Eq. (\ref{1.1}),
Eq. (\ref{jf39}) showing that this family only exists if $\varepsilon \gamma
<0$, i.e., exactly in the case opposite to that in which bright solitons are
found.

Out of the two families of the above soliton solutions (bright and dark) of
Eq. (\ref{1.1}), the bright family is the most interesting one. In spite of
their similarity to ordinary bright solitons, the bright soliton solutions
of Eq. (\ref{4aa.1}) feature a special property which distinguishes them
from ordinary solitary waves, namely, they are \textit{double-embedded
solitons}. The concept of embedded solitons ($ES$s) was formulated, in a
general form, in Ref. \cite{yang1}. It refers to solitary waves which do not
emit radiation, in spite of the fact that the soliton's wavenumber (spatial
frequency) is \textit{embedded} in the system's linear spectrum. Still
earlier, solitons of this type were found in particular models \cite{buryak}%
, for instance, in a generalized $NLS$ equation involving a quintic
nonlinear term \cite{fujioka}. Recently, more systems supporting $ES$s have
been found \cite{champneys1} -- \cite{Espinosa}. To the best of our
knowledge, the existence of $ES$s has not been reported before in models of $%
LC$ media.

So far, the embedded solitons were classified in two groups, namely, those
which obey $NLS$-like equations (or systems thereof), and those which are
governed by $KdV$-like equations. In the former case, an ES has its
wavenumber \textit{embedded} in the range of wavenumbers permitted to linear
waves (as it was already mentioned above). In the latter case, the velocity
of an ES is found in the range of phase velocities of linear waves. There
are, accordingly, two different ways to decide whether a solitary-wave
solution to a nonlinear $PDE$ system is embedded, viz., the \textit{%
wavenumber} (WN) and \textit{velocity} (VE) criteria.

In Ref. \cite{karpman} it was pointed out that Eq. (\ref{1.1}) is a
particular case of a more general $NLS$-like equation possessing $ES$s. For
this reason, and also in view of the significance of Eq. (\ref{1.1}) for
physical applications, it is interesting to determine if the bright-soliton
solutions of Eq. (\ref{1.1}) may be $ES$s. It should be noted that Eq. (\ref
{1.1}) may be regarded as both a $KdV$-like equation, due to its similarity
to the $mKdV$ one, \emph{and} an $NLS$-like equation, because, in the
context of wave propagation in $LC$s, Eq. (\ref{1.1}) in its complex form
plays a role similar to that of the $NLS$ equation, i.e., the one governing
evolution of a slowly varying envelope of a rapidly oscillating wave.
Therefore, it may be possible to apply \emph{both} criteria, WN and VE ones%
\textit{, }to decide if the soliton solutions of Eq. (\ref{1.1}) are $ES$s.

First, we apply the WN criterion. To this end, we must determine if the
wavenumber of the solution (\ref{4aa.4}) is contained within the range of
the wavenumbers allowed to linear waves. To identify the intrinsic
wavenumber of the solution, we must transform it into the reference frame
moving along the time axis with the reciprocal velocity $a$, see Eq. (\ref
{4aa.4}). The transformation adds a Doppler term to the soliton's internal
spatial frequency (wavenumber), making it equal to $q+ar$. On the other
hand, plane-wave solutions to the linearized version of Eq. (\ref{1.1}) in
the same reference frame can be sought for as 
\begin{equation}
u(z,t)=\exp \,i\left[ kz-\omega (t-az)\right] ,  \label{4aa.8}
\end{equation}
which leads to the following dispersion relation 
\begin{equation}
k(\omega )=\varepsilon \omega ^{3}-a\omega .  \label{4aa.9}
\end{equation}
Since the range in which the function (\ref{4aa.9}) takes its values covers
all the real numbers, including the soliton's wavenumber $q+ar$, all the
soliton solutions to\ Eq. (\ref{1.1}), given by Eqs. (\ref{4aa.4})-(\ref
{4aa.7}), are classified as $ES$s as per the $WN$ criterion.

Now, we address the question whether these solitons are also embedded
according to the $VE$ criterion. As the evolution variable in Eq. (\ref{1.1}%
) is the distance $z$, rather than the time $t$, it is the reciprocal
velocity which determines if the moving solutions are embedded according to
the $VE$ criterion. Thus, we should find out if the reciprocal velocity of
the soliton (\ref{4aa.4}), given by the parameter $a$, is contained within
the range of the reciprocal velocities permitted to linear waves. The
dispersion relation (\ref{4aa.9}) implies that the reciprocal phase
velocities of the linear waves (in the reference frame moving along with the
soliton) are given by 
\begin{equation}
\frac{k}{\omega }=-a+\varepsilon \omega ^{2},  \label{4aa.10}
\end{equation}
while the reciprocal velocity of the soliton proper is, obviously, zero in
the same reference frame. Obviously, the expression (\ref{4aa.10}) takes the
value zero if $a\varepsilon $ is positive, hence the soliton solutions given
by Eqs. (\ref{4aa.4}) -- (\ref{4aa.7}) are $ES$s according to the $VE$
criterion provided that $a\varepsilon >0$. As these solitons are also
embedded according to the $WN$ criterion, we call them \textit{%
double-embedded} solitons. On the other hand, when $a\varepsilon <0$, the
soliton solutions of Eq. (\ref{1.1}) are only embedded with respect to the $%
WN$ criterion, but not as per the $VE$ one, therefore in this case we apply
the term \textit{single-embedded} solitons.



\section{Radiation Inhibition and Continuity of the Embedded Solitons}

As in any other system with $ES$s, the fact that the solitons do not emit
radiation despite being embedded in the linear spectrum should be explained.
Since the wavenumber $q+ar$ of the soliton solution (\ref{4aa.4}) is
contained in the linear spectrum defined by the dispersion relation (\ref
{4aa.9}), a resonance of the soliton is expected with the linear waves whose
frequencies satisfy the condition 
\begin{equation}
q+ar=\varepsilon \omega ^{3}-a\omega .  \label{6.1}
\end{equation}
Moreover, when $a\varepsilon >0$ the soliton's reciprocal velocity $a$
coincides with the reciprocal phase velocities ($\varepsilon \omega ^{2}$)
of two linear waves whose frequencies satisfy the condition 
\begin{equation}
a=\varepsilon \omega ^{2},  \label{46a}
\end{equation}
consequently one could also expect the soliton to resonate with these
waves.\ Different explanations for the absence of resonant radiation in
other systems which support $ES$s were proposed \cite{champneys3}, \cite
{kosevich}. However, an explanation for the radiationless character of the $%
ES$s in Eq. (\ref{1.1}) has not been presented.

Another unexpected property of the same $ES$s in Eq. (\ref{1.1}) is the fact
that they exist in a continuous family. In most cases, $ES$s are isolated
solutions; usually they do not appear in families, although examples of
continuous families of $ES$s are known too, for instance, in a fifth-order $%
KdV$ equation \cite{yang2}. It is also necessary to explain why Eq. (\ref
{1.1}) has a two-parameter family of the $ES$ solutions.

As we show below, the radiationless character of the $ES$s in Eq. (\ref{1.1}%
) is the consequence of a special balance between the linear and the
nonlinear terms of this equation. To understand how these terms interact, it
will be helpful to separate their effects by considering the following
linear driven equation, 
\begin{equation}
\frac{\partial u}{\partial z}-\varepsilon \,\frac{\partial ^{3}u}{\partial
t^{3}}-\gamma \,\left| u_{0}\right| ^{2}\frac{\partial u_{0}}{\partial t}=0,
\label{6.2}
\end{equation}
where the source is built of a solution $u_{0}(z,t)$ to Eq. (\ref{1.1}). It
is clear that the same function $u_{0}$ is also a solution to Eq. (\ref{6.1}%
).

We now define the double Fourier transform of $u(z,t)$, 
\begin{equation}
\widetilde{u}(k,\omega )=\frac{1}{2\pi }\int\limits_{-\infty }^{\infty
}\int\limits_{-\infty }^{\infty }u(z,t)e^{-i(kz-\omega t)}dzdt,  \label{6.5}
\end{equation}
and Fourier transform Eq. (\ref{6.2}), to obtain 
\begin{equation}
\widetilde{u}(k,\omega )=i\frac{\widetilde{F}(k,\omega )}{-k+\varepsilon
\omega ^{3}},  \label{6.6}
\end{equation}
where 
\begin{equation}
F_{0}(z,t)=\gamma \,\left| u_{0}\right| ^{2}\frac{\partial u_{0}}{\partial t}
\label{6.4}
\end{equation}
is the source in Eq. (\ref{6.2}).

To understand the mechanism of the cancellation of the emission of
radiation, we can temporarily take, instead of the exact source (\ref{6.4}),
a model example with 
\begin{equation}
F_{0}(z,t)=A\,\mathrm{sech}\left( \frac{t-az}{w}\right) e^{i(qz+rt)}.
\label{6.7}
\end{equation}
In this case, the calculation of the Fourier transform of $F_{0}$ and
substitution in Eq. (\ref{6.6}) yield a result 
\begin{eqnarray}
\widetilde{u}(k,\omega ) &=&\frac{\pi wA\,\mathrm{sech}\left[ \frac{\pi }{2}%
w(r+\omega )\right] }{-(r+\omega )a-q+\epsilon \omega ^{3}}\left\{ -\frac{%
A^{2}\gamma r}{3}-\frac{w^{2}A^{2}\gamma r^{3}}{3}+\frac{A^{2}\gamma \omega 
}{6}\right.  \notag \\
&&\left. -\frac{w^{2}A^{2}\gamma r^{2}\omega }{2}+\frac{w^{2}A^{2}\gamma
\omega ^{3}}{6}\right\} \delta \left\{ \left[ (r+\omega )a+q\right]
-k\right\} .  \label{6.8}
\end{eqnarray}
At first sight, this expression seems to imply that a resonance with the
radiation waves should occur for frequencies at which the denominator, which
is a third-order polynomial in $\omega $, vanishes, which is actually
tantamount to Eq.(\ref{6.1}). Moreover, if $q=r=0$, the same argument shows
that a resonance at the frequencies defined by Eq. (\ref{46a}) should be
expected. Observe, however, that the numerator on the right-hand side of (%
\ref{6.8}) also contains a third-order polynomial in $\omega $.
Consequently, if the two polynomials happen to coincide, they will cancel
each other, which also implies the cancellation of the resonant generation
of the radiation modes. Equating the coefficients in front of powers of $%
\omega $ in the two polynomials in (\ref{6.8}), we obtain three equations
which, after some manipulations, take the \emph{precise} forms of Eqs. (\ref
{4aa.5}) -- (\ref{4aa.7}). Thus, these three equations are the necessary and
sufficient conditions for the mutual cancellation of the two polynomials in
Eq. (\ref{6.8}). This explains why the forcing term $F_{0}(z,t)$ of the form
(\ref{6.7}) does not generate any radiation, provided that the parameters $A$%
, $a$, $w$, $q$ and $r$ satisfy Eqs. (\ref{4aa.5}) -- (\ref{4aa.7}).
Furthermore, observe that the polynomial that appears in the numerator of \
the expression (\ref{6.8}) contains the nonlinear coefficient $\gamma $,
while the polynomial in the denominator contains the dispersion coefficient $%
\varepsilon $. Consequently, the cancellation between these two polynomials
is a result of the balance between the nonlinearity and dispersion in Eq. (%
\ref{1.1}). Finally, if the forcing term in Eq. (\ref{6.6}) is taken in the
exact form (\ref{6.4}), rather than in the simplified form of Eq. (\ref{6.7}%
), it is easy to check that the prediction for the cancellation of the
radiation emission will be the same.

\bigskip In the case of the full equation (\ref{1.1}), the same cancellation
argument explains why an initial condition of the form 
\begin{equation}
u(z=0,t)=A\,\mathrm{sech}\left( \frac{t}{w}\right) e^{irt}  \label{6.9}
\end{equation}
does not radiate at the frequencies defined by Eqs. (\ref{6.1})\ and (\ref
{46a}) if $A$ and $w$ do not satisfy Eq. (\ref{4aa.5}). On the other hand,
if $A$ and $w$ this condition, we expect the resonances to occur. In the
next section, we will verify numerically that this indeed the case.

To close this section, it is relevant to stress that the cancellation of the
two polynomials in Eq. (\ref{6.8}) imposes only three conditions, while the
solution (\ref{4aa.4}) involves five parameters. Therefore, the cancellation
conditions do not uniquely determine the soliton parameters, which explains
why the soliton solution (\ref{4aa.4})-(\ref{4aa.7}) involves two arbitrary
parameters, thus defining a two-parameter \emph{continuous} family of the $%
ES $s.

\section{Stability of the Embedded Solitons}

In this section we will study the stability of the bright-soliton solutions
of Eq. (\ref{1.1}). As it was explained in Sec. 4, these solitons may be
either \textit{single-embedded} or \textit{double-embedded}, depending on
the sign of the parameter combination $a\varepsilon $. In the following we
will separately consider the cases of positive and negative $a\varepsilon $.

We begin by considering a single-embedded soliton of Eq. (\ref{1.1}),
setting $\varepsilon =1$ and $\gamma =6$ [these values were chosen as they
correspond to those at which the related \textit{Hirota equation} \cite
{hirota}, which is connected to Eq. (\ref{1.1}) by the Galilean transform 
\cite{karpman}, is an exactly integrable one \cite{sasa}]. We start with the
following values of the soliton parameters 
\begin{equation}
A_{s}=\sqrt{\frac{5}{8}}\approx 0.790  \label{St1}
\end{equation}
\begin{equation}
w_{s}=\sqrt{\frac{8}{5}}  \label{St2}
\end{equation}
\begin{equation}
r_{s}=1/\sqrt{24}  \label{St3}
\end{equation}
\begin{equation}
a_{s}=-1/2  \label{St4}
\end{equation}
\begin{equation}
q_{s}=\frac{11}{6\sqrt{24}}.  \label{St5}
\end{equation}
These values satisfy the conditions (\ref{4aa.5}) -- (\ref{4aa.7}), and
therefore they characterize an exact bright soliton of the form (\ref{4aa.4}%
). Since $a_{s}\varepsilon <0$, this soliton is a single-embedded one (%
\textit{i.e.}, it is embedded solely according to the WN criterion).

To test stability of this soliton, we consider an initial condition of the
form 
\begin{equation}
u\left( z=0,t\right) =A_{0}\,\,\mathrm{sech}\mathit{\,}\left( \frac{t}{w_{0}}%
\right) \exp \,\left( ir_{0}t\right) ,  \label{St6}
\end{equation}
where $w_{0}=w_{s}$ and $r_{0}=r_{s}$, but $A_{0}$ is slightly different
from $A_{s}$. If we give $A_{0}$ a value $0.815$, which is larger than $A_{s}
$, the numerical solution of Eq. (\ref{1.1}) shows that the pulse moves to
the right along the temporal axis with a reciprocal velocity equal to $-0.58$%
, which is slightly lower than $a_{s}$, and the pulse's amplitude evolves as
shown in the upper curve of Fig. 2. The observation that the reciprocal
velocity of the\ perturbed pulse is lower than $a_{s}$ is consistent with
Eq. (\ref{4aa.6}), which indicates that $a$ should decrease if $A$ is
increased. Figure 2 (the upper curve) shows that the pulse's amplitude
stabilizes and approaches an equilibrium value close to $A=0.84$. The
temporal profile of the pulse at $z=50$ is displayed in Fig.\emph{\ }3. It
shows a small-amplitude radiation wave emitted by the trailing edge of the
pulse. The frequency composition of this tiny radiation wave can be
determined by calculating the Fourier transform ($FT$) of the radiation
contained in the interval $40\leq t\leq 168$. The insert in Fig. 3 shows the
power spectrum (\textit{i.e.}, the square of the $FT$ amplitude) of this
radiation, which contains two peaks located at the frequencies $\nu
_{1}=-0.10$\emph{\ }and\emph{\ }$\nu _{2}=0.12$. These peaks are close to
the resonant frequencies ($\nu =\pm 0.07$) predicted by the resonance
condition 
\begin{equation}
q_{s}+a_{s}r_{s}=\varepsilon \omega ^{3}-a_{s}\omega   \label{St7}
\end{equation}
[cf. Eq. (\ref{6.1})] and the \textit{partial resonance }condition \cite
{Espinosa} 
\begin{equation}
-\left( q_{s}+a_{s}r_{s}\right) =\varepsilon \omega ^{3}-a_{s}\omega .
\label{St8}
\end{equation}
These radiation peaks imply that the perturbed pulse emits radiation
according to the way the soliton's wavenumber is embedded in the spectrum of
linear waves.

If we now consider an initial condition of the form (\ref{St6}), with $%
w_{o}=w_{s}$, $r_{o}=r_{s}$ and $A_{o}=0.765<A_{s}$, the behavior of the
perturbed pulse is similar. In this case the amplitude evolves as shown in
the lower curve of Fig. 2 where we can see that the pulse's amplitude
approaches an equilibrium value close to $A=0.74$. The reciprocal velocity
of the perturbed pulse is $-0.42$, which is slightly higher than $a_{s}$.
This change is consistent with Eq. (\ref{4aa.6}), which indicates that $a$
should increase if $A$ is diminished.

The two curves shown in Fig. 2  demonstrate that the single-embedded soliton
solutions of Eq. (\ref{1.1}) are stable. This is an interesting result,
since usually $ES$s display a weak (nonlinear) one-sided instability \cite
{champneys3}. In fact, the complete stability of the $ES$s in Eq. (\ref{1.1}%
) may be expected, due to the fact that in this case we are dealing with a
continuous two-parameter family of the $ES$s, while in most other systems $ES
$s are isolated solutions, which explains their nonlinear instability.

Figure 2 also shows that if the amplitude of one of the single-embedded
solitons of Eq. (\ref{1.1}) is slightly increased, the perturbed soliton
stabilizes itself at an even higher amplitude. On the contrary, if the
soliton's amplitude is slightly decreased, the perturbed soliton stabilizes
at a still lower amplitude. This behavior \ can be better understood if we
analyze the evolution of the perturbed solitons of Eq. (\ref{1.1}) by means
of the averaged variational technique introduced by Anderson \cite{Anderson}%
, which is one of the approximately analytical methods used successfully in
nonlinear optics \cite{Lisak} -- \cite{Fujioka2}, see also a recent review 
\cite{progress}.

In order to apply the variational technique, we start with the \textit{ansatz%
} of the ordinary form, 
\begin{equation}
u(z,t)=A(z)\;\mathrm{sech}\left[ \frac{t-V(z)}{W(z)}\right] \;\exp \,i\left[
Q(z)+R(z)\,t+P(z)\,t^{2}\right] \emph{.}  \label{St9}
\end{equation}
Introducing this trial function in the Lagrangian density of Eq. (\ref{1.1}%
), 
\begin{equation}
L=i\,\left( u_{z}u^{\ast }-u_{z}^{\ast }u\right) +i\varepsilon \,\left(
u\,u_{ttt}^{\ast }-u^{\ast }u_{ttt}\right) +\frac{i\gamma }{2}\left[
u^{2}u^{\ast }u_{t}^{\ast }-(u^{\ast })^{2}u\,u_{t}\right] ,  \label{St10}
\end{equation}
and integrating over time, we calculate the averaged (effective) Lagrangian 
\begin{equation}
\mathcal{L}=\int_{-\infty }^{\infty }L\;dt\emph{.}  \label{St11}
\end{equation}
The following Euler-Lagrange equations can be easily derived from $\mathcal{L%
}$, 
\begin{equation}
-8AWQ^{\prime }-\frac{24\varepsilon AR}{W}-8\varepsilon AR^{3}W+\frac{16}{3}
\gamma A^{3}RW=f_{1}(P,P\,^{\prime },R\,^{\prime }),  \label{St12}
\end{equation}
\begin{equation}
-4A^{2}Q^{\prime }+\frac{12\varepsilon A^{2}R}{W^{2}}-4\varepsilon
A^{2}R^{3}+\frac{4}{3}\gamma A^{4}R=f_{2}(P,P\,^{\prime },A^{\prime
},V\,^{\prime },W\,^{\prime },R^{\prime }),  \label{St13}
\end{equation}
\begin{equation}
f_{3}(P,P\,^{\prime },A^{\prime },V\,^{\prime },W\,^{\prime },R\,^{\prime
})=0,  \label{St14}
\end{equation}
\begin{equation}
A^{2}W=A^{2}(0)\,W(0),  \label{St15}
\end{equation}
\begin{equation}
-\frac{12\varepsilon A^{2}}{W}-12\varepsilon A^{2}R^{2}W+\frac{4}{3}\gamma
A^{4}W=f_{4}(P,P\,^{\prime },A^{\prime },V\,^{\prime },W\,^{\prime
},R\,^{\prime }),  \label{St16}
\end{equation}
\begin{equation}
-\frac{24\varepsilon A^{2}V}{W}-24\varepsilon A^{2}R^{2}VW+\frac{8}{3}\gamma
A^{4}VW=f_{5}(P,P\,^{\prime },A^{\prime },V\,^{\prime },W\,^{\prime
},R\,^{\prime }),  \label{St17}
\end{equation}
where the primes stand for the \textit{z}-derivatives, and the expressions $%
f_{n}(P,P\,^{\prime },A^{\prime },V\,^{\prime }$ $,W\,^{\prime },R\,^{\prime
})$ are nonlinear functions of their arguments. Their explicit forms are not
given, as they will not be needed in what follows.

We now resort to search for fixed points of Eqs. (\ref{St12}) -- (\ref{St17}%
), which are stationary solutions of the form 
\begin{equation}
A^{\prime }=W\,^{\prime }=R\,^{\prime }=P\,^{\prime }=P=0,  \label{St18}
\end{equation}
\begin{equation}
Q\,^{\prime }=\mathrm{const}\equiv q,  \label{St19}
\end{equation}
\begin{equation}
V\,^{\prime }=\mathrm{const}\equiv a.  \label{St20}
\end{equation}
When we insert these conditions into Eqs. (\ref{St12})-(\ref{St17}), we find
that\ $f_{n}=0$ (for $n=1,...,5$) and consequently, the following relations
are obtained 
\begin{equation}
A^{2}W\,^{2}=\frac{18\,\varepsilon }{\gamma },  \label{St21}
\end{equation}

\begin{equation}
a=3\,\varepsilon R\,^{2}-\frac{1}{6}\gamma A^{2},  \label{St22}
\end{equation}

\begin{equation}
q=\frac{1}{2}\gamma A^{2}R-\varepsilon R\,^{3},  \label{St23}
\end{equation}
\begin{equation}
A^{2}W=\mathrm{const}=A^{2}(0)\,W(0).  \label{St24}
\end{equation}
Equation (\ref{St21}) is the variational counterpart of Eq. (\ref{4aa.5}),
and the expressions for $a$ and $q$ coincide exactly with those in (\ref
{4aa.6}) and (\ref{4aa.7}). On the other hand, Eq. (\ref{St24}) applies not
only to stationary solutions, but to general dynamical equations as well,
with variable $A(z)$ and $W(z)$, as it expresses the variational version of
the exact conservation law (which is simply the energy conservation in the
case of nonlinear optics \cite{progress}).

Equation (\ref{St24}) is plotted by thin curves, corresponding to two
different initial conditions, in Fig. 4. This figure also shows plots (the
bold curve) of Eq. (\ref{St21}), corresponding to $\varepsilon =1$ and $%
\gamma =6$. This diagram helps to understand why the soliton [characterized
by the parameters (\ref{St1}) -- (\ref{St5}) and marked by point \textbf{E}
on the bold curve in Fig. 4], if perturbed by increasing or decreasing its
initial amplitude, stabilizes itself, as was observed in Fig. 2.

We take, as the initial perturbed soliton, the one corresponding to point 
\textbf{1} in Fig. 4.\textbf{\ }It has the same width as the unperturbed
soliton at point \textbf{E}, but a larger amplitude, 
\begin{equation}
A_{1}=0.815>0.790=A_{E}.  \label{St25}
\end{equation}
According to Eq. (\ref{St24}), the perturbed pulse must evolve sliding along
the thin curve passing through point \textbf{1. }The thin curve intersects
the equilibrium bold curve at point \textbf{2}, which is therefore a fixed
point.\textbf{\ }Within the framework of the variational approximation
proper, the trajectory may perform some oscillations in a vicinity of this
fixed point; however, if effective loss due to the emission of small amounts
of radiation by the perturbed soliton (which was observed above in direct
simulations) is taken into regard\textbf{,} the trajectory will be attracted
to the fixed point, and will eventually end up being trapped at this point,
thus implying the stabilization of the soliton very close to point \textbf{%
2, }which has the value of the amplitude $0.840$.

Similarly, starting at the initial condition corresponding to point \textbf{3%
}, the soliton will slide along the thin line until it gets stuck at the
stable fixed point \textbf{4. }As the amplitude corresponding to point 
\textbf{4 }is $0.740$, the origin of the stabilization process observed in
direct simulations displayed in the lower curve of Fig. 2 is now clear.

So far we considered relaxation of perturbed single-embedded soliton. Now we
proceed to the stability of double-embedded ones. To this end, we set $%
\varepsilon =\gamma =1$, and choose the soliton parameters 
\begin{equation}
A_{d}=\sqrt{\frac{5}{8}}\approx 0.790,  \label{St26}
\end{equation}
\begin{equation}
w_{d}=\sqrt{\frac{48}{5}},  \label{St27}
\end{equation}
\begin{equation}
r_{d}=1/4,  \label{St28}
\end{equation}
\begin{equation}
a_{d}=1/12\approx 0.08,  \label{St29}
\end{equation}
\begin{equation}
q_{d}=1/16.  \label{St30}
\end{equation}
These values satisfy the conditions (\ref{4aa.5}) -- (\ref{4aa.7}),
therefore they define an exact bright soliton of the form (\ref{4aa.4}).
Since $a_{d}\varepsilon >0$, this soliton is a double-embedded one. We
perturb it by taking an initial condition of the form (\ref{St6}) with $%
w_{0}=w_{d},$ $r_{0}=r_{d}$, and $A_{0}=0.815>A_{d}$.

The numerical solution of Eq. (\ref{1.1}) corresponding to this initial
condition shows that the perturbed pulse moves along the temporal axis with
a reciprocal velocity equal to $0.07$, which is slightly lower than $a_{d}$
(this lower velocity is consistent with Eq. (\ref{4aa.6})). Simultaneously,
the pulse's amplitude oscillates as shown in the upper curve of Fig. 5. This
figure again shows a trend of the perturbed pulse to stabilize. However, in
this case (with the double-embedded soliton) the stabilization process is
slower, and it is necessary to pass a greater distance (along the $z$ axis)
to observe the damping of the amplitude oscillations. The upper curve of
Fig. 5  shows that the pulse's amplitude eventually approaches an
equilibrium value close to $0.84$.

The trailing edge of the perturbed double-embedded soliton emits a tiny
radiation wavetrain whose frequency components can be determined by
calculating the $FT$ of the radiation contained in the interval $45\leq
t\leq 109$ (for $z=200$). The spectrum obtained in this way is shown in Fig.
6. In this figure two peaks are seen. The bigger one corresponds to the
frequency $\nu =-0.046\,$ ($\omega =-0.289$), which corresponds to the
negative solution of Eq. ( \ref{46a}), and therefore it is a consequence of
the resonance of the perturbed soliton with a linear wave whose phase
velocity is equal to the soliton's velocity. On the other hand, the smaller
radiation peak is located at $\nu =0.078$ ($\omega =0.490$), which is very
near to the only real root ($\omega =1/2$) of the resonance condition 
\begin{equation}
q_{d}+a_{d}\,r_{d}=\varepsilon \omega ^{3}-a_{d}\,\omega .  \label{St31}
\end{equation}
Therefore, the latter peak is due to the fact that the soliton's wavenumber $%
q_{d}+a_{d}\,r_{d}$ is contained in the range of wavenumbers permitted to
linear waves.

As the larger radiation peak (the one at $\nu =-0.046$) exists due to the
fact that the soliton is embedded according to the VE criterion, one could
assume that in this case (\textit{i.e.,} when a double-embedded soliton is
perturbed) the radiation emitted by the pulse is mainly due to the VE
embedding of the soliton. However, such a conclusion would be wrong. The
left radiation peak in Fig. 6 actually has a larger amplitude because the $FT
$ of the complete solution  is slightly shifted to the left (as a
consequence of $r_{0}$ being positive), and it is this shift which enhances
the left radiation peak.

To verify the latter point, one can consider a slightly different initial
condition, characterized by the parameters $A_{0}=0.815$, $w_{0}=w_{d}=\sqrt{%
48/5}$, and $r_{0}=-r_{d}=-1/4$. As in this case $r_{0}$ is negative, the $FT
$ of the complete solution will be shifted to the right, and this shift will
enhance the right radiation peak. At the insert in Fig. 6 we show the
spectrum of the radiation emitted in this case by the perturbed
double-embedded soliton (for $z=100$). As expected, in this case the
radiation peak due to the wavenumber embedding of the soliton (\textit{i.e.,}
the right peak) is much higher than the one existing due to the velocity
embedding (the very small peak on the left). We thus conclude that both
embeddings, WN and VE, are important to explain the emission of radiation by
perturbed double-embedded solitons.

If we now consider an initial pulse of the form (\ref{St6}) with $w_{0}=w_{d}
$, $\,r_{0}=r_{d}$ and $A_{0}=0.765<A_{d}$, the numerical solution of Eq. (%
\ref{1.1}) shows that the pulse's amplitude again performs a damped
oscillatory behavior, as shown in the lower curve of Fig. 5. In this case,
the pulse's amplitude approaches an equilibrium value close to $0.74$. 

\section{Concluding Remarks}

In this work, using the multiple scales method, we have derived a model for
the propagation of a wavepacket of $TM$ modes along a cylindrical
liquid-crystal waveguide beyond the usual weakly nonlinear limit of the Kerr
medium. In this case, the amplitude of the wavepacket obeys a nonlinear
equation, (\ref{1.1}) or (\ref{4aa.1}), which exhibits a derivative
nonlinearity. This complex modified $KdV$ equation gives rise to the
two-parameter families of bright, Eqs. (\ref{4aa.4}) -- (\ref{4aa.7}), and
dark, Eqs. (\ref{jf38}) -- (\ref{jf41}), solitons. The bright-soliton
solutions of Eq. (\ref{1.1}) are embedded solitons ($ES$s) (or sometimes
double-embedded ones), \textit{i.e.}, they do not emit any radiation, in
spite of the fact that their wavenumbers (and sometimes their velocities
too) fall into the linear spectrum of the system. We have shown that the
physical nature of the existence of the $ES$s inside the continuous spectrum
is the balance between the dispersion and nonlinearity in Eq. (\ref{1.1}).
Moreover, it was concluded that these $ES$s are completely stable solutions,
while, in most previously considered models, they are weakly unstable. It
was observed that perturbed single-embedded solitons relax to a new
equilibrium state faster than double-embedded ones.

The coupled expansions for $\theta $ and $H_{\phi }$ in powers of $q$, that
were introduced in Sec. 2, can be extended to higher orders. This leads to
nonlinear equations with the quintic i.e., $O(q^{5})$, nonlinearity.
Investigation of the corresponding model is currently in progress. Also, as
discussed in Sec. 3, up to the order $O(q^{4})$ considered here, the same
procedure to construct narrower ($\alpha =2$) or wider ($\alpha =1/2$)
wavepackets of $TM$ modes can also be carried out.

Another possible generalization of our model, not dealt with here, is a
possibility to take into account hydrodynamic flows beyond the Kerr-medium
approximation, that will inevitably couple to the reorientation dynamics of
the liquid crystal. Actually, the inclusion of the flow is unavoidable owing
to the fluid nature of the system. However, the consideration of the
hydrodynamical part of the system substantially complicates the problem.
Some effects produced by this generalization were considered, at the level
of the $NLS$ approximation, \textit{i.e.}, at order $O(q^{3})$, in Ref. \cite
{Rodriguez1prima}.

{\Large Acknowledgments}

We acknowledge a partial financial support from DGAPA-UNAM IN105797, from
FENOMEC through the grant CONACYT 400316-5-G25427E and from CONACYT 41035,
M\'{e}xico. We also thank DGSCA-UNAM (Direcci\'{o}n General de Servicios de
C\'{o}mputo Acad\'{e}mico de la UNAM) for their authorization to use the
computer Origin 2000 during this work.

{\LARGE Figure Captions}

Fig. 1. Schematic of a laser beam propagating through the nematic
liquid-crystal cylindrical guide. Transverse-magnetic (TM) modes are shown
explicitly.

Fig. 2 Evolution of the amplitude of two perturbed single-embedded solitons
of Eq. (\ref{1.1}) (with $\varepsilon =1$ and $\gamma =6$). The upper curve
corresponds to the initial condition (\ref{St6}) with $A_{0}=0.815>A_{s}$, $%
w_{0}=w_{s}$ and $r_{0}=r_{s}$, where $A_{s}$, $w_{s}$ and $r_{s}$ are the
values (\ref{St1}) -- (\ref{St3}). The lower curve corresponds to a similar
initial condition with $A_{0}=0.765<A_{s}$ [$A(z)$ and $z$ are dimensionless
quantities].

Fig. 3. Temporal profile (at $z=50$) of the perturbed single-embedded
soliton of Eq. (\ref{1.1}) whose amplitude (as a function of $z$ is shown in
the upper curve of Fig. 2. The spectrum of this profile is shown in the
insert [$u$ and $t$ are dimensionless quantities].

Fig. 4. The bold curve passing through the point $E=(A_{E},W_{E})=(0.790,%
\,2.192)$ is the plot of Eq. (\ref{St21}) with $\varepsilon =1$ and $\gamma
=6$. The thin line passing through point 1 plots Eq. (\ref{St24}) with $%
A(0)=0.815>A_{E}$ and $W(0)=W_{E}$. The thin line passing through point 3 is
also a plot of Eq. (\ref{St24}), with $A(0)=0.765<A_{E}$ and $W(0)=W_{E}$.

Fig. 5. Evolution of the amplitude of two perturbed double-embedded solitons
of Eq. (\ref{1.1}) (with $\varepsilon =\gamma =1$). The upper curve
corresponds to the initial condition (\ref{St6}) with $A_{0}=0.815>A_{d}$, $%
w_{0}=w_{d}$, and $r_{0}=r_{d}$ [where $A_{d}$, $w_{d}$ and $r_{d}$ are
given by Eqs. (\ref{St26}) -- (\ref{St28})], and the lower curve corresponds
to a similar initial condition with $A_{0}=0.765<A_{d}$ [$A(z)$ and $z$ are
dimensionless quantities].

Fig. 6. Spectrum (obtained at $z=200$) of  the radiation emitted by the
perturbed double-embedded soliton whose amplitude (as a function of $z$) is
shown in the upper curve of Fig. 5. The insert shows the spectrum (obtained
at $z=100$) of the radiation emited when the sign of $r_{d}$ is reversed (%
\textit{i.e., }when\textit{\ }$r_{d}=-1/4$).

\newpage

$\bigskip $\appendix 

{\Large Appendix A}

Expressions for the coefficients $A_{0}$, $A_{1}$, $A_{2}$, $A_{3}$, $A_{4}$%
, $B_{0}$, $B_{1}$, $C_{0}$, $C_{1}$, $C_{2}$, $C_{3}$ and $C_{4}$, which
appear in Eq. (\ref{2.16}):

\begin{equation}
A_{0}(a,b;\gamma ,\epsilon _{\perp })=4a^{3}b(a+b)\gamma \epsilon _{\perp },
\label{A.1}
\end{equation}
\begin{eqnarray}
A_{1}(a,b;\beta ,\gamma ,\mu ,\epsilon _{a},\epsilon _{\perp },k_{0})
&=&a^{2}\{-24ab(a+b)[\beta ^{2}\epsilon _{a}-\epsilon _{\perp }(\epsilon
_{a}+\epsilon _{\perp })\mu k_{0}^{2}]+  \notag \\
&&\gamma \lbrack 16a(3a-b)b^{2}\beta ^{2}\epsilon
_{a}-2(a^{2}+ab+8b^{2})\epsilon _{\perp }+  \notag \\
&&16ab^{2}(-3a+b)\epsilon _{\perp }(\epsilon _{a}+\epsilon _{\perp })\mu
k_{0}^{2}]-  \notag \\
&&ab(a+b)\epsilon _{\perp }\gamma ^{2}\},  \label{A.2}
\end{eqnarray}
\begin{equation}
A_{2}(a,b;\beta ,\gamma ,\mu ,\epsilon _{a},\epsilon _{\perp
},k_{0})=-12ab(a+b)\gamma \lbrack -\epsilon _{\perp }+4a^{2}(\beta
^{2}\epsilon _{a}-\epsilon _{\perp }(\epsilon _{a}+\epsilon _{\perp })\mu
k_{0}^{2})],  \label{A.3}
\end{equation}
\begin{eqnarray}
A_{3}(a,b;\beta ,\gamma ,\mu ,\epsilon _{a},\epsilon _{\perp },k_{0})
&=&a\{24b(a+b)(\beta ^{2}\epsilon _{a}-\epsilon _{\perp }(\epsilon
_{a}+\epsilon _{\perp })\mu k_{0}^{2})+  \notag \\
&&2\gamma \lbrack 32a^{2}b\beta ^{2}\epsilon _{a}+8ab^{2}\beta ^{2}\epsilon
_{a}+8b^{3}\beta ^{2}\epsilon _{a}+  \notag \\
&&a\epsilon _{\perp }-7b\epsilon _{\perp }-8b(4a^{2}+ab+b^{2})\epsilon
_{\perp }(\epsilon _{a}+\epsilon _{\perp })\mu k_{0}^{2}]+  \notag \\
&&\gamma ^{2}b(a+b)\epsilon _{\perp }\},  \label{A.4}
\end{eqnarray}
\begin{equation}
A_{4}(a,b;\beta ,\gamma ,\mu ,\epsilon _{a},\epsilon _{\perp
},k_{0})=16ab(a+b)\gamma \lbrack -\beta ^{2}\epsilon _{a}+\epsilon _{\perp
}(\epsilon _{a}+\epsilon _{\perp })\mu k_{0}^{2}],  \label{A.5}
\end{equation}
\begin{equation}
B_{0}(a,b;\gamma ,\epsilon _{\perp })=-2ab(a-b)(a+b)^{2}\gamma \epsilon
_{\perp },  \label{A.6}
\end{equation}
\begin{equation}
B_{1}(a,b;\beta ,\gamma ,\mu ,\epsilon _{a},\epsilon _{\perp
},k_{0})=ab(a-b)(a+b)^{2}[24\beta ^{2}\epsilon _{a}-24\epsilon _{\perp
}(\epsilon _{a}+\epsilon _{\perp })\mu k_{0}^{2}+\gamma ^{2}\epsilon _{\perp
}],  \label{A.7}
\end{equation}
\begin{equation}
C_{0}(a,b;\gamma ,\epsilon _{\perp })=4ab^{3}(a+b)\gamma \epsilon _{\perp },
\label{A.8}
\end{equation}
\begin{eqnarray}
C_{1}(a,b;\beta ,\gamma ,\mu ,\epsilon _{a},\epsilon _{\perp },k_{0})
&=&b^{2}\{24ab(a+b)[\beta ^{2}\epsilon _{a}-\epsilon _{\perp }(\epsilon
_{a}+\epsilon _{\perp })\mu k_{0}^{2}]+  \notag \\
&&\gamma \lbrack -16a^{2}b(a-3b)\beta ^{2}\epsilon
_{a}-2(8a^{2}+3ab+3b^{2})\epsilon _{\perp }  \notag \\
&&+16a^{2}b(a-3b)\epsilon _{\perp }(\epsilon _{a}+\epsilon _{\perp })\mu
k_{0}^{2}  \notag \\
&&+ab(a+b)\gamma \epsilon _{\perp }\},  \label{A.9}
\end{eqnarray}
\begin{equation}
C_{2}(a,b;\beta ,\gamma ,\mu ,\epsilon _{a},\epsilon _{\perp
},k_{0})=-12ab(a+b)\gamma \lbrack -\epsilon _{\perp }+4b^{2}(\beta
^{2}\epsilon _{a}-\epsilon _{\perp }(\epsilon _{a}+\epsilon _{\perp })\mu
k_{0}^{2})],  \label{A.10}
\end{equation}
\begin{eqnarray}
C_{3}(a,b;\beta ,\gamma ,\mu ,\epsilon _{a},\epsilon _{\perp },k_{0})
&=&b\{24a(a+b)[-\beta ^{2}\epsilon _{a}+\epsilon _{\perp }(\epsilon
_{a}+\epsilon _{\perp })\mu k_{0}^{2}]+  \notag \\
&&2\gamma (8a^{3}\beta ^{2}\epsilon _{a}+8a^{2}b\beta ^{2}\epsilon
_{a}+32ab^{2}\beta ^{2}\epsilon _{a}-5a\epsilon _{\perp }  \notag \\
&&+3b\epsilon _{\perp }-8a(a^{2}+ab+4b^{2})\epsilon _{\perp }(\epsilon
_{a}+\epsilon _{\perp })\mu k_{0}^{2})  \notag \\
&&-a(a+b)\epsilon _{\perp }\gamma ^{2}\},  \label{A.11}
\end{eqnarray}
\begin{equation}
C_{4}(a,b;\beta ,\gamma ,\mu ,\epsilon _{a},\epsilon _{\perp
},k_{0})=16ab(a+b)\gamma \lbrack -\beta ^{2}\epsilon _{a}+\epsilon _{\perp
}(\epsilon _{a}+\epsilon _{\perp })\mu k_{0}^{2}]\,.
\end{equation}

An explicit form of $\theta ^{(2)}(x)$:

\begin{eqnarray}
\theta ^{(2)}(x) &=&\frac{\beta \epsilon _{a}e^{-(b+ax)\gamma }J_{1}^{2} [%
\sqrt{\epsilon _{c}(\frac{\omega _{0}a}{c})^{2}-\beta ^{2}a^{2}]}} {%
24x^{2}a^{2}b(a+b)^{2}(b-a)\pi ^{3}\gamma ^{2}\epsilon _{\perp }^{3}\epsilon
_{\parallel }^{2}}\left[ -a^{2}e^{(1+x)a\gamma }\left( 1-x\right) \left(
a^{2}\left\{ 24bax\beta ^{2}\epsilon _{a}\right. \right. \right.  \notag \\
&&-48bx\beta ^{2}\gamma \epsilon _{a}(b-ax)+2(xa-2b)\gamma \epsilon _{\perp
}+bax\epsilon _{\perp }\gamma ^{2}+24bxa(\frac{\omega _{0}a}{c})^{2}  \notag
\\
&&\left. (-1+2b\gamma -2xa\gamma \epsilon _{\perp }\epsilon _{\parallel
})\right\} +bxa\left\{ 8bax\beta ^{2}\epsilon _{a}(3+2b\gamma -2xa\gamma
)\right.  \notag \\
&&\left. +2\epsilon _{\perp }\gamma (xa-2b+\gamma bax/2)-8bxa(3+2b\gamma
-2xa\gamma )\epsilon _{\perp }\epsilon _{\parallel }(\frac{\omega _{0}a}{c}
)^{2}\right\}  \notag \\
&&+ab(a-b)(a+b)^{2}e^{-(b+a)\gamma }\left[ -2\gamma \epsilon _{\perp
}+xa\left( 24\beta ^{2}\epsilon _{a}\right. \right.  \notag \\
&&\left. \left. +\epsilon _{\perp }(\gamma ^{2}-24\epsilon _{\parallel } (%
\frac{\omega _{0}a}{c})^{2})\right) \right] +b(xa-b)e^{(b+ax)\gamma }\left(
6bax\gamma \epsilon _{\perp }(b+xa)\right.  \notag \\
&&\left. +16a^{4}x(b+xa)\gamma \left[ \beta ^{2}\epsilon _{a}-\epsilon
_{\parallel }(\frac{\omega _{0}a}{c})^{2}\right] \right) +a^{2}\left(
-8\epsilon _{a}xa\beta ^{2}\left[ 3b+6b^{2}\gamma \right. \right.  \notag \\
&&\left. +xa(3+2xa\gamma )\right] -\gamma \epsilon _{\perp }\left[
4(b-3xa)+xa\gamma (b+xa)\right] +  \notag \\
&&\left. +8xa\left[ 3b+6b^{2}\gamma +xa(3+2xa\gamma )\right] \epsilon
_{\perp }\epsilon _{\parallel }(\frac{\omega _{0}a}{c})^{2}\right) +a\left(
-10x^{2}a^{2}\gamma \epsilon _{\perp }\right.  \notag \\
&&+4b^{2}\left[ -6xa\beta ^{2}\epsilon _{a}+12x^{2}a^{2}\beta ^{2}\gamma
\epsilon _{a}-\gamma \epsilon _{\perp }-\gamma ^{2}xa\epsilon _{\perp
}/4\right.  \notag \\
&&\left. -6xa(2xa\gamma -1)\epsilon _{\perp }\epsilon _{\parallel }(\frac{
\omega _{0}a}{c})^{2}\right] +4bxa\left[ -6xa\beta ^{2}\epsilon
_{a}-4x^{2}a^{2}\beta ^{2}\gamma \epsilon _{a}\right.  \notag \\
&&\left. \left. \left. +\gamma \epsilon _{\perp }/2-\gamma ^{2}xa\epsilon
_{\perp }/4+2xa(2xa\gamma +3)\epsilon _{\perp }\epsilon _{\parallel }(\frac{
\omega _{0}a}{c})^{2}\right] \right) \right] \,.  \label{A.13}
\end{eqnarray}

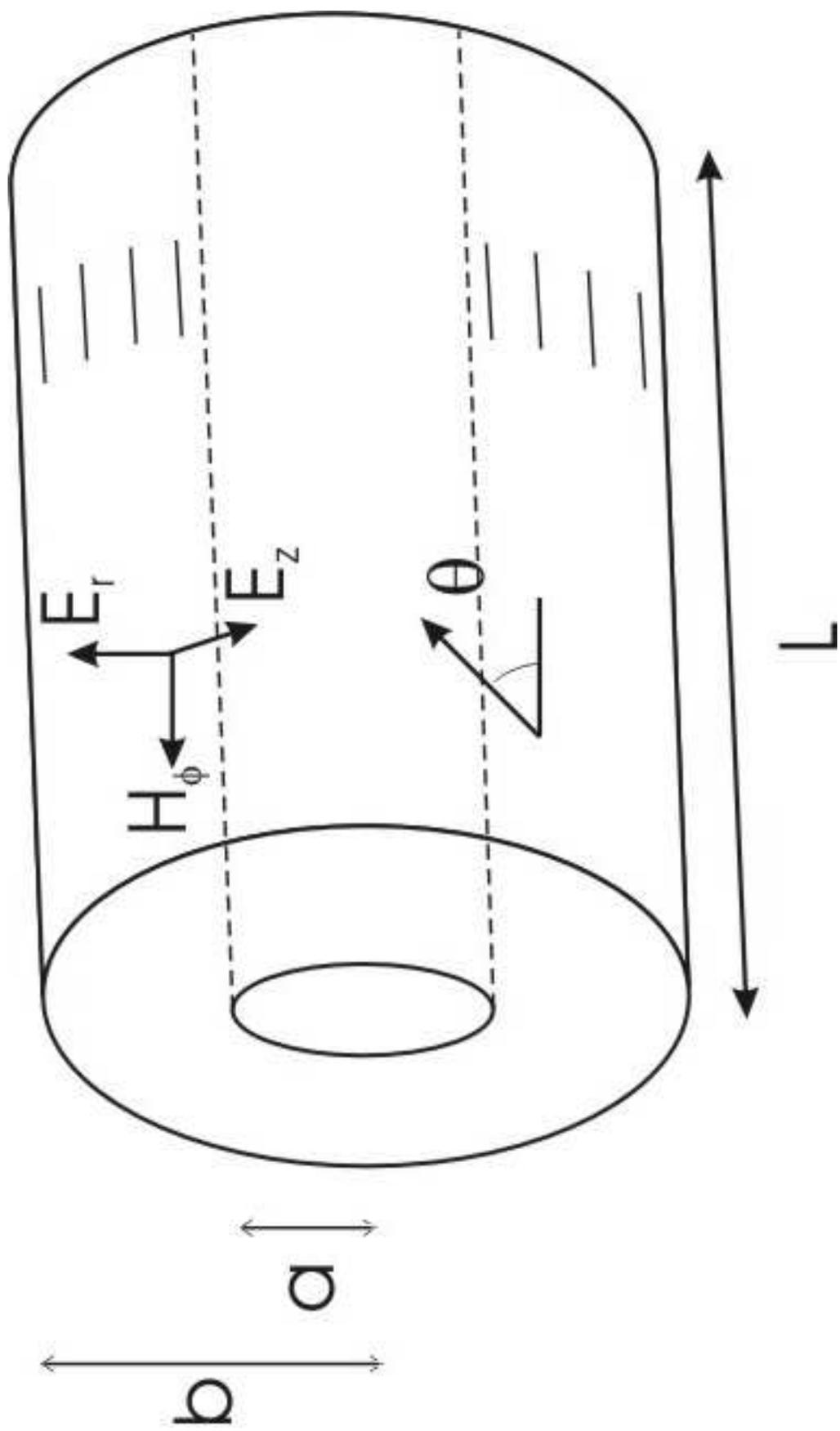
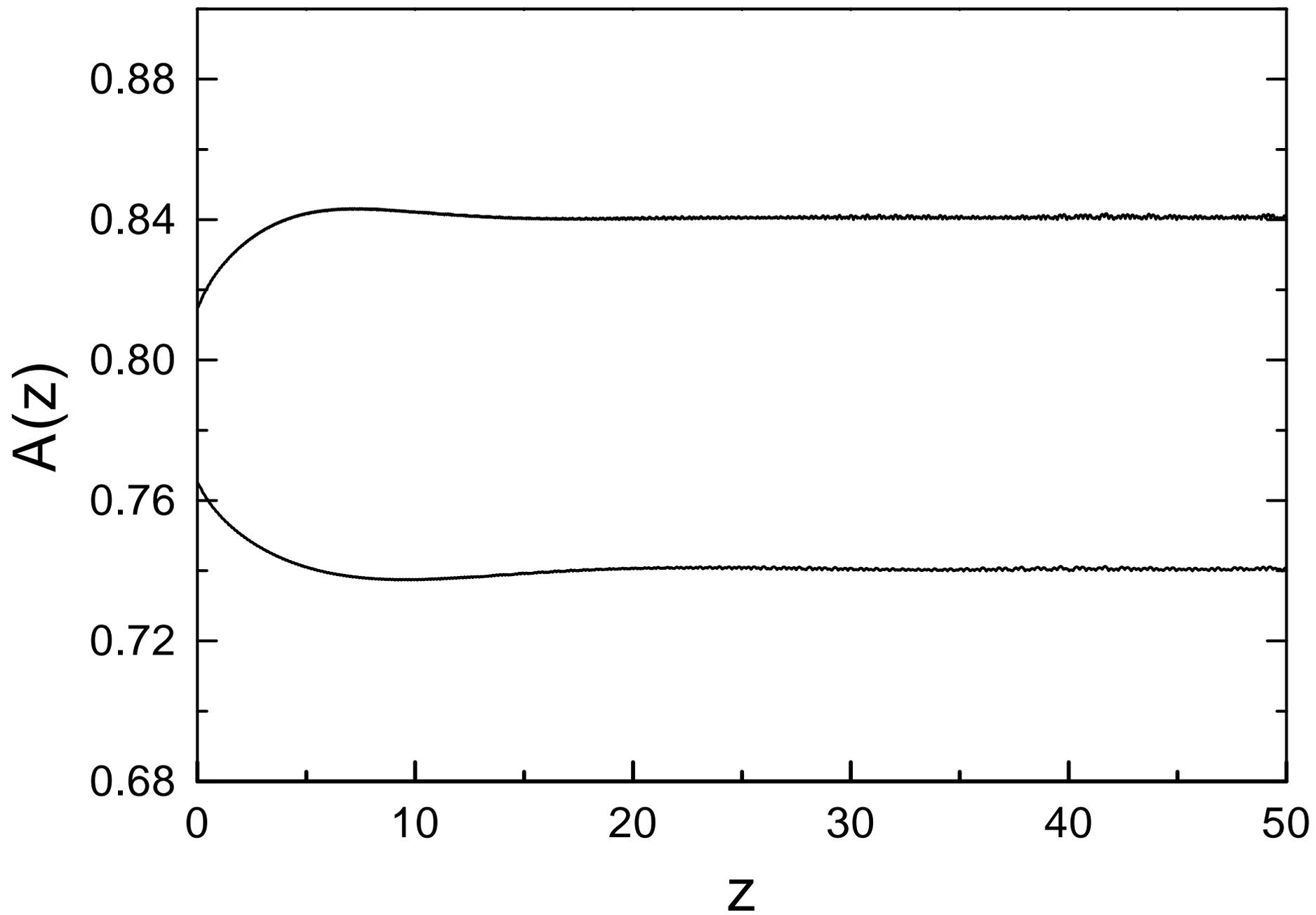
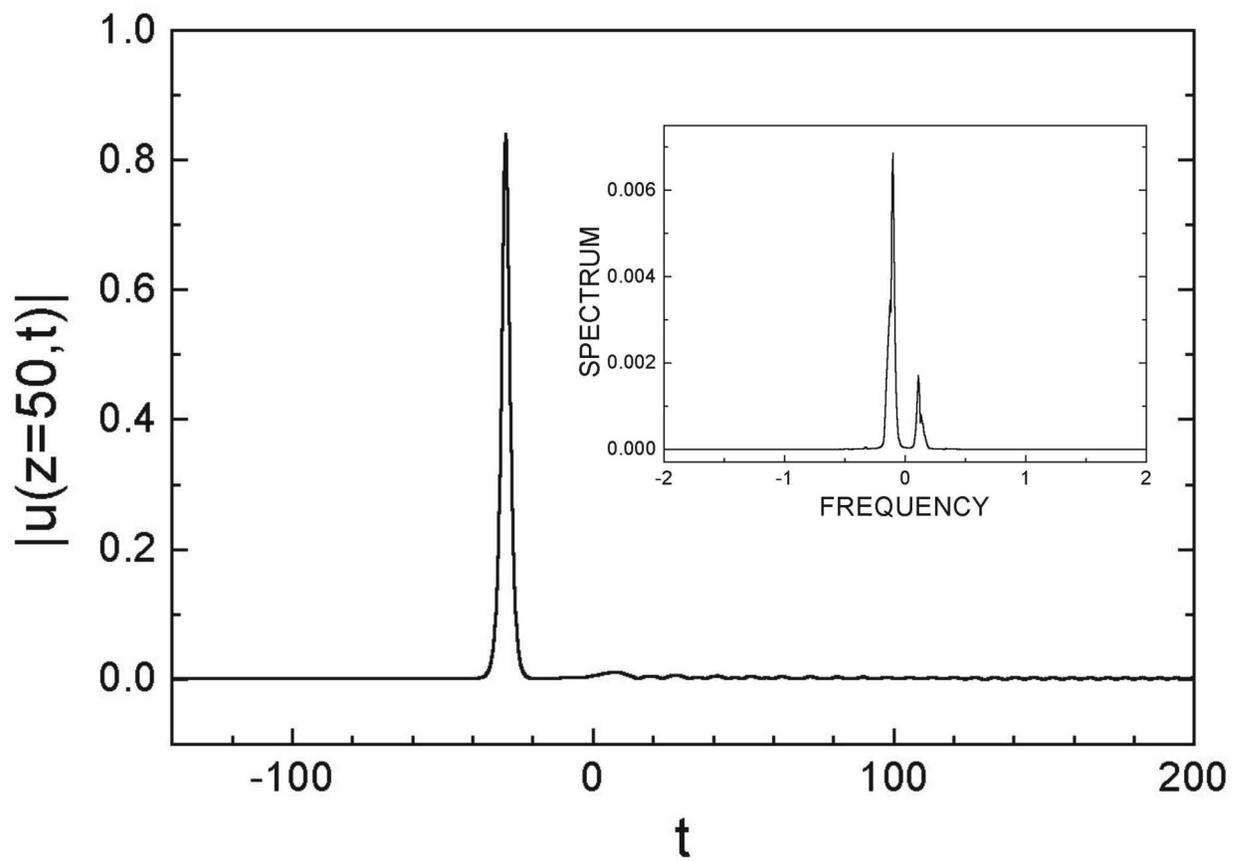
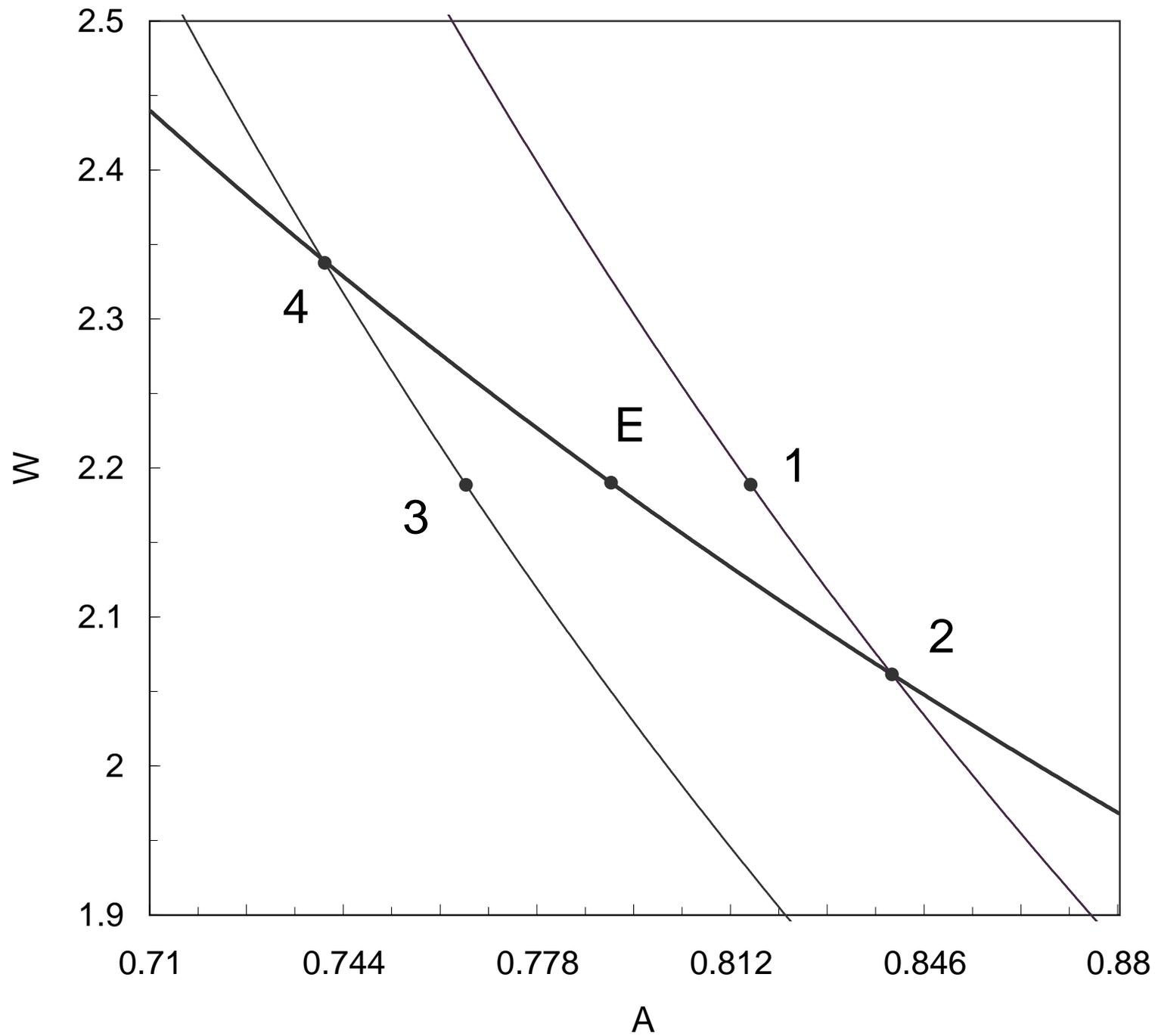
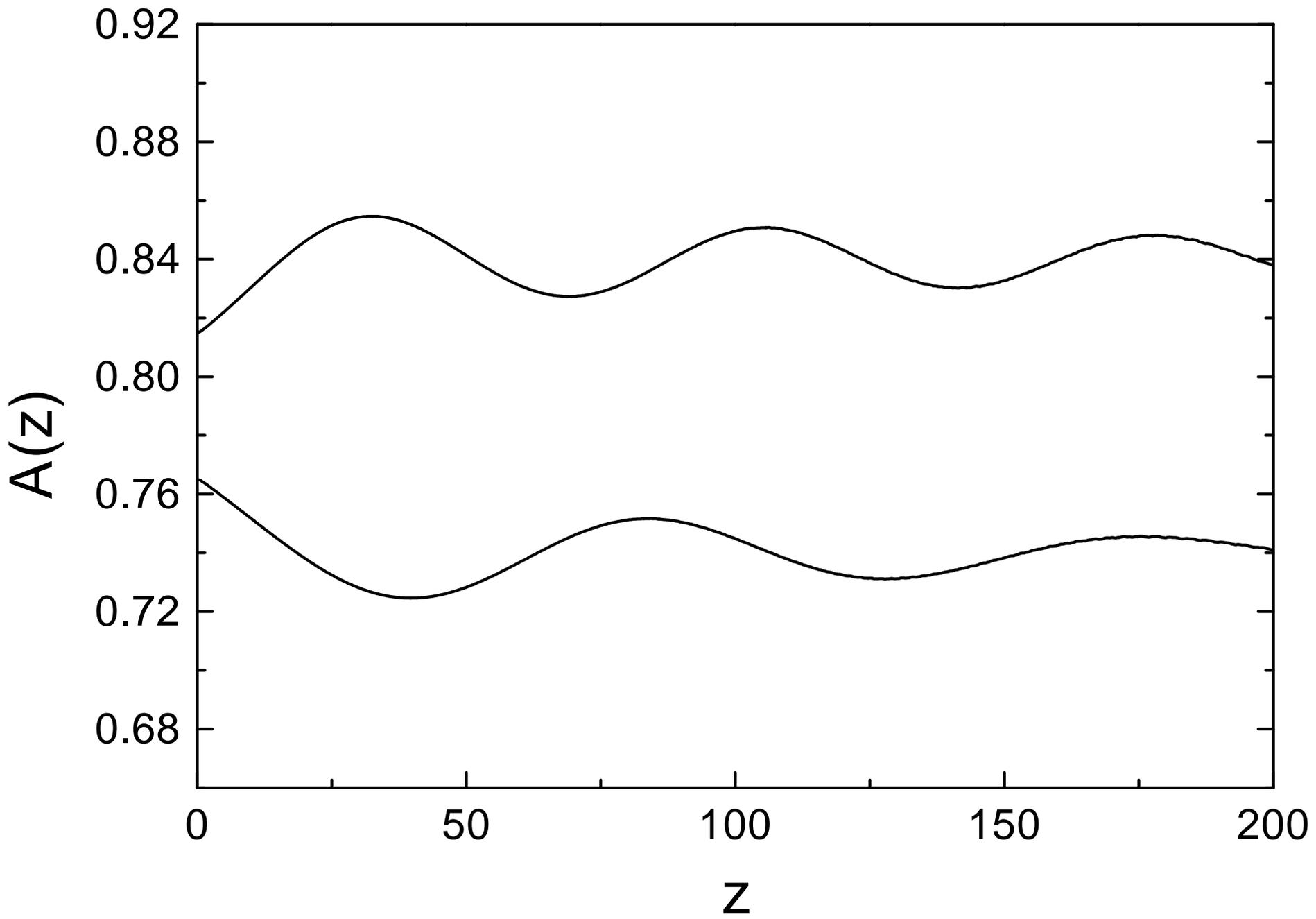
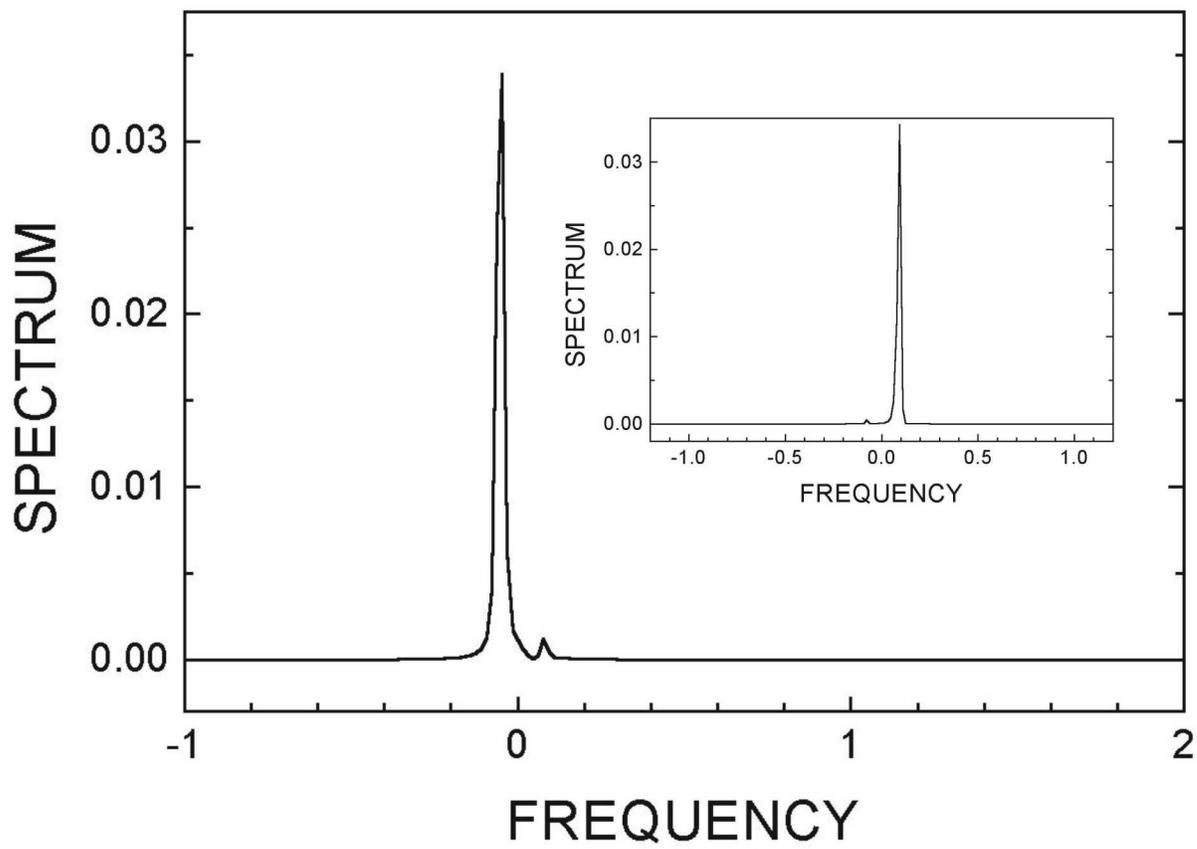
\end{document}